\documentstyle{elsart}
\input psfig.sty

\begin{document}
\begin{frontmatter}
\title{Triplet Pairing in Neutron Matter}

\author{V. V. Khodel},
\author{V. A. Khodel}, and
\author{J. W. Clark}
\address{McDonnell Center for the Space Sciences
and Department of Physics, \\
Washington University, St. Louis, MO 63130-4899}

\begin{abstract}
The separation method developed earlier by us [Nucl.\ Phys.\ {\bf A598} 390 
(1996)] to calculate and analyze solutions of the BCS gap equation for 
$^1$S$_0$ pairing is extended and applied to $^3$P$_2$--$^3$F$_2$ pairing in
pure neutron matter.  The pairing matrix elements are written as a separable 
part plus a remainder that vanishes when either momentum variable is on the 
Fermi surface.  This decomposition effects a separation of the problem 
of determining the dependence of the gap components in a spin-angle 
representation on the magnitude of the momentum (described by a set of
functions independent of magnetic quantum number) from the problem of 
determining the dependence of the gap on angle or magnetic projection.  
The former problem is solved through a set of nonsingular, quasilinear 
integral equations, providing inputs for solution of the latter problem 
through a coupled system of algebraic equations for a set of numerical 
coefficients.  An incisive criterion is given for finding the upper 
critical density for closure of the triplet gap.  The separation method 
and its development for triplet pairing exploit the existence of a 
small parameter, given by a gap-amplitude measure divided by the Fermi 
energy.  The revised BCS equations admit analysis revealing universal 
properties of the full set of solutions for $^3$P$_2$ pairing in the 
absence of tensor coupling, referring especially to the energy degeneracy 
and energetic order of these solutions.  The angle-average approximation 
introduced by Baldo et al.\ is illuminated in terms of the 
separation-transformed BCS problem and the small parameter expansion.  
Numerical calculations of $^3$P$_2$ pairing parameters and gap functions, 
with and without coupling to the $^3$F$_2$ state, are carried out for 
pairing matrix elements supplied by (vacuum) two-neutron interactions 
that fit nucleon-nucleon scattering data.  It is emphasized that {\it ab 
initio} evaluation of the in-medium particle-particle interaction and 
associated single-particle energies will be required if a reliable 
quantitative description of nucleonic superfluids is to be achieved.
\end{abstract}
\end{frontmatter}

\section{Introduction \label{sec:intro}}
The study of superfluidity in infinitely extended nuclear systems has a 
long history [1--82], predating the 1967 discovery of pulsars \cite{hewish}, 
which were soon identified as rapidly rotating magnetic neutron 
stars \cite{gold}.  Interest in nucleonic pairing has intensified 
in recent years, owing primarily to experimental developments on two 
different fronts.  In the field of astrophysics, a series of $X$-ray 
satellites (including Einstein, EXOSAT, ROSAT, and ASCA) has brought a 
flow of data on thermal emission from neutron stars, comprising both upper 
limits and actual flux measurements.  The recent launching of the 
Chandra $X$-ray observatory provides further impetus for more incisive 
theoretical investigations.  Realistic {\it ab initio} prediction of the 
microscopic physics of nucleonic superfluid components in the interiors 
of neutron stars is crucial to a quantitative understanding of neutrino 
cooling mechanisms [85--95] that operate immediately after their birth 
in supernova events, as well as the magnetic properties, vortex structure, 
rotational dynamics, and pulse timing irregularities [96--99] of these 
superdense stellar objects.  In particular, when nucleonic species enter 
a superfluid state in one or another region of the star, suppression 
factors of the form $\exp (-\Delta_F /k_B T)$ are introduced 
into the expression for the emissivity, $\Delta_F$ being an appropriate 
average measure of the energy gap at the Fermi surface.  On the terrestrial 
front, the expanding capabilities of radioactive-beam and heavy-ion 
facilities have stimulated a concerted exploration of exotic nuclei and 
nuclei far from stability, with special focus on neutron-rich species 
\cite{riisager,mueller}.  Pairing plays a prominent role in modeling 
the structure and behavior of these new nuclei.  

In this article we will be principally concerned with nucleonic pairing
in the astrophysical setting.  To a first approximation, a neutron star 
is described as a neutral system of nucleons (and possibly heavier baryons) 
and electrons (and possibly muons) in beta equilibrium at zero temperature, 
with a central density several times the saturation density $\rho_0$ 
of symmetrical nuclear matter \cite{pethick,shapiro,lamb,fiks,alparlives}. 
The gross structure of the star (mass, radius, pressure and density 
profiles) is determined by the Tolman-Oppenheimer-Volkov general 
relativistic equation of hydrostatic equilibrium, consistently with 
the continuity equation and the equation of state (which embodies the 
microscopic physics of the system).  The star contains (i) an {\it outer 
crust} made up of bare nuclei arranged in a lattice interpenetrated 
by relativistic electrons, (ii) an {\it inner crust} where a similar 
Coulomb lattice of neutron-rich nuclei is embedded in Fermi seas of 
relativistic electrons and neutrons, (iii) a {\it quantum fluid interior} 
of coexisting neutron, proton, and electron fluids, and finally 
(iv) a {\it core region} of uncertain constitution and phase (but 
possibly containing hyperons, a pion or kaon condensate and/or quark matter).

It is generally accepted \cite{ccdk,ttr} that the neutrons in the 
background fluid of the inner crust of a neutron star will pair-condense 
in the $^1$S$_0$ state.  Qualitatively, this phenomenon can be understood
as follows.  At the relatively large average particle spacing at the 
``low'' densities involved in this region ($\rho \sim \rho_0/10$), 
the delocalized neutrons experience mainly the attractive component 
the $^1$S$_0$ interaction.   However, the pairing effect is quenched 
at higher densities, $\sim \rho_0$ and beyond, due to the strong 
short-range component of this interaction.  By similar reasoning, one 
expects $^1$S$_0$ proton pairing to occur in the quantum fluid interior, 
in a density regime where the proton contaminant (necessary for
charge balance and chemical equilibrium) reaches a partial 
density $\rho_p \sim \rho_0/10$.  

The energy dependence of the nucleon-nucleon ($NN$) phase shifts in different 
partial waves offers some guidance in judging what nucleonic pair-condensed 
states are possible or likely in different regions of a neutron star.
A rough correspondence between baryon density and $NN$ bombardment energies
can be established through the Fermi momenta assigned to the nucleonic
components of neutron-star matter \cite{ttr}, which provide a crude
measure of the maximum relative momenta attained in the nuclear medium.
In (pure) neutron matter, only $T=1$ partial waves are allowed by the
Pauli principle.  Moreover, one need only consider partial 
waves with $L\leq 4$ in the range of baryon density -- optimistically,
$\rho < (3-4) \rho_0$ -- where a nucleonic model of neutron-star material 
is tenable.  The $^1$S$_0$ phase shift is positive at low energy 
(indicating an attractive in-medium force) but turns negative (repulsive) 
at around 250 MeV lab energy.  Thus, unless the in-medium pairing force 
is dramatically different from its vacuum counterpart, the situation 
already suggested above should prevail:  S-wave pairs should form at 
low densities but should be inhibited from forming when the density 
approaches that of ordinary nuclear matter.  

The next lowest $T=1$ partial waves are the three triplet 
P waves $^3$P$_J$, with $J=0,1,2$, which are actually coupled 
by the tensor force to the triplet F waves with the same total 
angular momentum $J$.  Of these channels, the first two are not
good candidates for pairing.  In the $^3$P$_0$ state, the
phase shift is only mildly attractive at low energy, turning
repulsive at a lab energy of 200 MeV, while the $^3$P$_1$ phase
shift is repulsive at all energies.  On the other hand, the
$^3$P$_2$ phase shift is negative at all energies, and indeed
is the most attractive $T=1$ phase shift at energies above
about 160 MeV.  This exceptional attractive strength in the
$^3$P$_2$ state is due mainly to the spin-orbit interaction and in
part to the coupling to $^3$F$_2$ wave via the tensor component
of the $NN$ force.  A substantial pairing effect in the
$^3$P$_2$--$^3$F$_2$ channel may be expected at densities somewhat 
in excess of $\rho_0$, again assuming that the relevant in-vacuum 
interaction is not greatly altered within the medium.  

The remaining $T=0$ partial waves with $L\leq 4$ are both
singlets: $^1$D$_2$ and $^1$G$_4$.  However, the $^1$D$_2$
phase shift, though positive over the energy domain of interest,
is clearly dominated by the $^3$P$_2$ phase shift, while the
$^1$G$_4$ phase shift, again positive, is even less important.

Therefore, in the conventional picture of neutron-star matter (see, 
however, ref.~\cite{alm3}), singlet and triplet pairing are operationally 
equivalent to $^1$S$_0$ and $^3$P$_2$--$^3$F$_2$ pairing, respectively.  
There have been many illuminating studies of the former problem, notably 
\cite{chen,ccdk,baldo1,kkc,jensen}.  Here we shall concentrate on triplet 
pairing in the pure neutron system, ignoring the relatively minor 
modifications that arise in beta-stable neutral matter with its dilute 
admixture of protons.

BCS theory \cite{bcs,schrieffer} furnishes a general framework for 
microscopic treatments of nucleonic pairing.  As is the case for singlet 
pairing \cite{ccdk,kkc}, two rather different problems must be overcome 
to realize a quantitative description of triplet pairing within
this framework and make realistic predictions for the associated 
energy gap.  The first problem to be solved is the construction of an 
accurate in-medium (or ``effective'') particle-particle interaction 
that provides the pairing force, together with a consistent mass operator 
that provides the single-particle input for the BCS equation.  
The second problem is to solve the BCS gap equation (actually
a system of gap equations when one deals with $J\neq 0$), once
the inputs for the pairing matrix and normal-state single-particle
energies have been determined.

This first task places great demands on even the most advanced tools
of many-body theory.  For $^1$S$_0$ pairing, there exists a ``first 
generation'' of results on this facet of the problem, obtained by 
several different approaches (Landau Fermi-liquid theory \cite{kallman},
the method of correlated basis functions \cite{chen,ccdk}, the
polarization-potential approach \cite{pines,wambach} and diagrammatic 
perturbation theory \cite{schulze2}).  However, little progress has been 
made on this aspect of $^3$P$_2$--$^3$F$_2$ pairing, and the issue of 
modifications of the associated input pairing interaction and 
single-particle energies due to short- and long-range correlations 
inside the nuclear medium remains open.  We shall not attempt to address
this issue here.  Rather, we shall be concerned with the task
of actually solving the triplet BCS system, assuming the particle-particle
interaction is identical with the bare or vacuum $NN$ interaction,
and that the input single-particle energies needed to complete
the BCS equation(s) have the same form as for free particles
(perhaps with an effective mass).  This problem also proves to
be nontrivial, since we face essentially the same difficulties
as in singlet S-wave pairing, and more:  
\begin{enumerate}
\item[(a)]
Again, the kernel of the BCS gap equation(s) has a pole at zero gap 
amplitude, implying an incipient logarithmic divergence.  
\item[(b)]
Again, the nonlocality of the particle-particle interaction in momentum space 
implies slow convergence of the integral over momentum modulus $k$ in the 
BCS equation(s) -- precluding application of the standard weak-coupling
estimates of BCS theory \cite{schrieffer}.  This feature of nucleonic 
pairing implies that one must integrate out to large values of the 
momentum variable.
\item[(c)] 
As already intimated, an additional complication of pairing in 
higher angular momentum states with $J\neq 0$ arises from
the (possible) angle-dependence of the gap, which can be
parametrized in terms of dependences on the quantum numbers
$J,L$ of the total and relative orbital angular momenta and
on the magnetic quantum number $M$ associated with $J$.  One 
must deal with a set of coupled gap equations for the gap components 
$\Delta_L^{JM}(k)$ in an expansion of the (generally anisotropic) 
gap in the spin-angle functions ${\hat G}_L^{JM}({\bf n})$.  
\end{enumerate}

The combination of difficulties (a) and (b) can make numerical
solution of the gap equation(s) a problematic exercise. To overcome 
these obstacles, a special procedure called the separation approach 
has been introduced in ref.~\cite{kkc} and applied to $^1$S$_0$ pairing. 
In this approach, the generic pairing matrix $V(k,k')$ is decomposed, 
identically, into a separable part, plus a remainder whose every
element vanishes when either momentum variable lies on the Fermi 
surface.  (As is characteristic of BCS theory, we assume that
the particle-particle interaction is independent of frequency.) 
Applied to $^1$S$_0$ pairing, this separation procedure reduces
the BCS gap equation to two equivalent coupled equations:  a
nonsingular, quasilinear integral equation for the shape of
the gap function and a nonlinear equation determining its
amplitude at the Fermi surface.  We hasten to point out that
this strategy is by no means limited to the case where the
pairing matrix elements are supplied by the bare $NN$ 
interaction -- it will retain its practical and analytical
value when a realistic in-medium particle-particle interaction
is employed.

The primary aim of this article is to generalize the separation
approach to pairing in an arbitrary angular momentum channel
and apply it to $^3$P$_2$ pairing in neutron matter.  Since,
according to consideration (c) above, we are now confronted --
in general -- with a set of coupled gap equations rather than
a single one, the situation is much more complicated than for $^1$S$_0$ 
pairing.  However, in this case the power of the separation
method becomes even more apparent.  It enables a decomposition
of the problem into (i) the determination of the shape of the gap
function in $k$-space through an appropriate nonsingular quasilinear 
integral equation whose kernel is practically independent of the energy
gap, and, once this is done, (ii) the determination of a set of 
numerical coefficients specifying the angular dependence of the 
gap through a system of nonlinear `algebraic' equations.  

In both singlet and triplet cases, the effectiveness of the separation 
scheme is greatly enhanced by the existence of a small parameter, 
given by the ratio of a suitable measure of the gap amplitude to 
the Fermi energy.  Small-parameter expansions work well in the
singlet problem, and even better in the triplet case.  Corrections to 
the leading order in a small-parameter expansion of the 
$^3$P$_2$--$^3$F$_2$ energy gap are down by a factor $10^{-5}$ or 
$10^{-6}$.   This property leads to considerable simplifications in 
the analytical and numerical treatment of the triplet problem, greatly 
ameliorating complication (c).   

The separation approach is not merely a tool for numerical solution
of gap equations.  Following the lead of refs.~\cite{kkc,luso,univ,van}, 
we shall demonstrate its very important role in facilitating incisive 
analysis of the nature of pairing solutions in different angular momentum 
channels.  In particular, the separation formalism provides a basis 
for establishing certain universal properties of $^3$P$_2$ pairing, 
properties that are independent of density and independent of the 
specifics of the pairing interaction itself, whether furnished by the 
in-vacuum or in-medium particle-particle interaction.  The analytical 
results obtained here and in refs.~\cite{univ,van} remain valid 
outside the context of neutron matter and carry implications for the 
still more complex problem of pairing in liquid $^3$He \cite{vol}.

In sect.~\ref{sec:gapeqn}, we display the BCS gap equations that describe
pairing in an arbitrary two-body channel of given total spin $S$ and 
isospin $T$.  Sect.~\ref{sec:sepformalism} extends the separation method of 
ref.~\cite{kkc} to this general problem, with special attention to 
$^3$P$_2$--$^3$F$_2$ pairing.  Sect.~\ref{sec:nocoupling} is devoted to an 
analysis of $^3$P$_2$ pairing in the absence of tensor coupling to the 
$^3$F$_2$ state (``pure'' $^3$P$_2$ pairing).  We offer a straightforward
criterion for determination of the upper critical density at which a 
solution of the pairing problem ceases to exist in the given two-body 
channel, in terms of vanishing of a characteristic determinant associated
associated with the separation formulation.  In sect.~\ref{sec:aaa} we 
introduce the small-parameter expansion of the pairing problem and use 
it to clarify the nature of the angle-average approximation proposed by 
Baldo et~al.~\cite{baldo2}.  In turn, the latter approximation is adopted
to formulate an efficient procedure that allows reasonably accurate 
numerical solution of the coupled-channel, $^3$P$_2$--$^3$F$_2$ pairing 
system when the goal is to find the magnitude of the pairing effect and 
the detailed structure of the pairing solutions is a lesser concern.  We 
go on to recount how the fundamental character of the diverse solutions of 
the nonlinear pairing problem may be elucidated within the separation 
method.  The derivation of universal features of $^3$P$_2$ pairing is 
exemplified and discussed. Results from numerical application of the 
separation approach to triplet pairing in neutron matter are reported
in sect.~\ref{sec:tripletnumapp}.  Three representative models of the 
free-space $NN$ interaction are used to construct pairing matrix 
elements, the Argonne $v_{18}$ potential \cite{argonne18} being chosen 
as an example from the current generation of models yielding high-precision 
fits of the $NN$ scattering data \cite{nijmegenp}.  The results presented 
explicitly include the critical upper density for gap closure, the momentum 
dependence of gap components, and the conventional average measure of 
the gap amplitude at the Fermi surface, variously calculated for 
``pure'' and tensor-coupled $^3$P$_2$ pairing, with and without use of 
the angle-average approximation.  These results are compared with previous 
findings of other authors.  In sect.~\ref{sec:concl}, we indicate some 
promising future directions for research on pairing in infinitely 
extended nucleonic systems.  An appendix reviews the separation method 
as developed for $^1$S$_0$ pairing.  Small-parameter expansion is employed 
to devise an approximation for the gap amplitude that may be taken 
as a generalization of the BCS weak-coupling formula.  Additionally, 
the technique of small-parameter expansion is extended to the triplet 
pairing problem.

\section{Any-channel gap equation \label{sec:gapeqn}}
To describe pairing in an arbitrary two-body state with total spin $S$,
angular momentum $J$, and isospin $T$, it is useful to introduce a
$2 \times 2$ gap matrix ${\hat \Delta}({\bf k})$ which has the 
expansion (cf.~\cite{ttr,t72})
\begin{equation}
{\hat \Delta}({\bf k}) = \sum_{LJM} \Delta^{JM}_L(k)
{\hat G}^{JM}_L({\bf n})
 \label{matgap}
\end{equation}
in the spin-angle matrices
\begin{equation}
{\hat G}^{JM}_L({\bf n})  \equiv\sum_{M_S M_L} \langle \frac12\frac12
\sigma_1\sigma_2| S M_S\rangle \langle SL M_S M_L |
JM\rangle Y_{LM_L} ({\bf n}) \, ,
\label{spinm}
\end{equation}
with ${\bf n} \equiv {\bf k}/k$.
The total spin and isospin quantum numbers $S,T$ ($=(1,1)$ for triplet
states and neutron matter) are fixed throughout and generally suppressed.
Setting $S^{JMJ_1M_1}_{L'L_1}({\bf n})={\rm Tr}
\left[{\hat G}^{JM*}_{L'}({\bf n}){\hat G}_{L_1}^{J_1M_1}({\bf n})\right]$,
the spin-angle matrices (\ref{spinm}) may be shown to obey the orthogonality
condition
\begin{equation}
\int S^{JMJ_1M_1}_{L'L_1}({\bf n}){\rm d}{\bf n}=\delta_{L'L_1}
\delta_{JJ_1}\delta_{MM_1} \, .
\label{ort1}
\end{equation}
The particle-particle interaction has the corresponding expansion
\begin{equation}
V({\bf k},{\bf k}')=\sum_{LL'M}\langle k|V_{LL'}^J|
k' \rangle {\hat G}^{JM}_L({\bf n}){\hat G}_{L'}^{JM*}({\bf n}') \, ,
\label{potex}
\end{equation}
with $|L-L'|\leq 2$ in the case of tensor forces.  The coupled set
of generalized gap equations for the components $\Delta^{JM}_L(k)$
then reads
\begin{equation}
\Delta_L^{JM}(k)= \sum_{L'L_1J_1M_1}(-1)^{\Lambda} \int
  \langle k|V_{LJ}^{L'J}|k'\rangle S^{JMJ_1M_1}_{L'L_1}({\bf n}')
  {\Delta_{L_1}^{J_1M_1}(k')\over 2E({\bf k}')}{\rm d}{\bf n}'
  {\rm d}\tau' \, ,
\label{gengap}
\end{equation}
with $\Lambda \equiv S+(L-L')/2$ and 
$ {\rm d}\tau' \equiv (2/\pi)(k')^2{\rm d} k' $.  The energy 
denominator $E({\bf k})=[\xi^2(k)+D^2({\bf k})]^{1/2}$ 
involves the square of the single-particle excitation energy 
$\xi(k)=\varepsilon(k)-\mu$ of the normal 
system, measured relative the Fermi energy $\mu$, and a term
\begin{equation}
 D^2({\bf k})={1\over 2}\sum_{LJML_1J_1M_1}\Delta_L^{JM*}(k)
 \Delta_{L_1}^{J_1M_1}(k) S^{JMJ_1M_1}_{LL_1}({\bf n})
\label{squd}
\end{equation}
giving rise to an energy gap in the single-particle excitation
spectrum of the pair-condensed system.  Examination of the
the explicit formula for the angular factor
$S^{JMJ_1M_1}_{LL_1}({\bf n})$ reveals
that a summation over total momentum $J_1$ must be present in (\ref{gengap})
and (\ref{squd}). (This fact is well known in the theory of superfluid
$^3$He \cite{vol}: the anisotropic ABM phase is built of states with
total momentum $J=1$ {\it and} 2.)  The sum over orbital momenta $L'$ on
the l.h.s. of (\ref{gengap}) appears only when tensor forces
are present and is then restricted by $|L-L'|\leq 2$ and parity
conservation.  On the other hand, the sum over $L_1$ resulting from 
the angular dependence of the gap contains, in principle, an infinite 
number of terms.

We identify the energy gap as $|D({\bf k})|$ and note that
it is in general angle dependent and in general has nodes.  It is
convenient to define an angle-averaged gap function $\Delta(k)$
through the positive root of
\begin{equation}
\Delta^2(k) \equiv {\overline{D^2}}(k)
= \frac1{4\pi} \int  D^2({\bf k})d{\bf n} = 
{1 \over {4\pi}} {1\over 2} \sum_{LJM} \left| \Delta_L^{JM}(k)\right|^2
\, .
\label{aagap}
\end{equation}
(The last equality follows from the orthogonality property (3).)  
Analogously to the $^1$S$_0$ pairing problem, the quantity (7),
evaluated at the Fermi momentum $k_F$ and denoted
$\Delta_F^2\equiv {\overline{D^2}}(k_F)$
furnishes an overall measure of the strength of pairing
correction to the ground-state energy in the preferred state.
The notation ${\overline{D^2}}(k_F)$ is included here since
it has been used earlier in the literature \cite{ttr}.

We take this opportunity to caution the reader that the literature
is inconsistent and sometimes vague in the choice of energy gap measure.  
Care must be taken in interpreting statements about the magnitude of 
``the gap.'' Moreover, the choice made in normalizing or defining the 
pairing matrix elements is not uniform.  We have attempted to be precise,
and for the most part we follow the conventions established by Takatsuka
and Tamagaki \cite{tt71,ttr}.  

\section{Separation formulation of the triplet pairing 
problem\label{sec:sepformalism}}

Generalizing the separation technique developed in ref.~\cite{kkc}
for the $^1$S$_0$ problem, we again split the pairing
matrix elements into a separable part and a remainder $ W^J_{LL'}(k,k')$
that vanishes when either of the momentum variables $k$, $k'$ lies on the
Fermi surface.  Thus
\begin{equation}
\langle k'| V_{LL'}^J| k \rangle \equiv V_{LL'}^J(k,k') =
  v_{LL'}^J\phi_{LL'}^J(k)\phi_{LL'}^J(k')
 + W_{LL'}^J(k,k') \, ,
\end{equation}
where $v_{LL'}^J \equiv V_{LL'}^J (k_F,k_F) \equiv
\langle k_F | V^J_{L'L} | k_F \rangle$ and
$\phi_{LL'}^J= \langle k_F | V_{LL'}^J | k_F \rangle  / v_{LL'}^J$.
Substitution of this identity into the generic-channel gap
equation (\ref{gengap}) yields
\begin{eqnarray}
\Delta^{JM}_L(k)&-&\sum_{L'} (-1)^\Lambda \int  W_{LL'}^J (k,k')
\sum_{L_1J_1M_1}S_{L'L_1}^{JMJ_1M_1} ({\bf n'})
{\Delta^{J_1M_1}_{L_1}(k')\over 2\sqrt{\xi^2(k')
+ \delta^2}}{\rm d}{\bf n}' {\rm d} \tau' \nonumber \\
&=&  \sum_{L'} D^{JM}_{LL'}\phi_{LL'}^J(k)
\label{delsf}
\end{eqnarray}
where the coefficients
\begin{equation}
D^{JM}_{ LL'} =  (-1)^\Lambda v_{LL'}^J\int \phi^J_{LL'}(k)
\sum_{L_1J_1M_1} S^{JMJ_1M_1}_{L'L_1}({\bf n}) \frac{\Delta^{J_1M_1}_{L_1}(k)}
{2\sqrt{\xi^2(k) + D^2({\bf k})}}{\rm d} {\bf n} {\rm d} \tau
\label{defdf}
\end{equation}
(with ${\rm d} \tau \equiv (2/\pi) k^2 dk)$ are just numbers.

As in ref.~\cite{kkc} (see also the appendix), we have introduced
a typical scale $\delta$ in place of the gap in single-particle
spectrum, in the term of (\ref{delsf}) involving the portion
$W_{LL'}^J (k,k')$ of the pairing matrix that vanishes on the
Fermi surface.  Such terms are quite insensitive to the choice of
this scale, since the vanishing of $W_{LL'}^J (k,k')$ at $k'=k_F$ 
ensures that the integral over $k'$ receives its overwhelming
contributions some distance from the Fermi surface where
$D^2(k')$ is negligible compared to $\xi^2(k')$.
In the present context where the gap functions are angle-dependent,
it is important to note that the $\delta$ quantity may be regarded
as angle-independent. Numerical calculations show that the final
result is practically independent of the choice of $\delta$
and we shall normally take it to be zero.

Employing the orthogonality condition (\ref{ort1}), one finds
that only diagonal terms (meaning $L_1=L'$, $J_1=J$, $M'=M$)
contribute to the sum on the l.h.s.\ of eq.~(\ref{delsf}), which
becomes
\begin{equation}
\Delta_L^{JM}(k) - \sum_{L'} (-1)^\Lambda \int
W_{LL'}^J(k,k')\frac{\Delta_{L'}^{JM}(k') } {2|\xi(k')|}{\rm d}
\tau'= \sum_{L'} D_{LL'}^{JM}\phi_{LL'}^J(k) \,.
\label{base1}
\end{equation}
Henceforth we shall frequently suppress the fixed index $J$. The
functions $\phi_{LL_1}^J(k)\equiv \phi^{LL_1}(k)$ superposed on
the r.h.s.\ of (\ref{base1}) do not depend on $M$.  Accordingly,
the gap components $\Delta_L^M(k)=\Delta_L^{JM}(k)$ can be constructed
as linear combinations
\begin{equation}
\Delta^M_L(k)=\sum_{L_1L_2} D^M_{L_1L_2}\chi^{L_1L_2}_L(k)
\label{gen1}
\end{equation}
of $M-$independent functions $ \chi^{L_1L_2}_L(k)= 
\chi^{JL_1L_2}_L(k)$ multiplied by
coefficients $D^M_{L_1L_2}$ determined by eq.~(\ref{defdf}).
It is helpful to regard the quantities $\Delta^M_L(k)$ as forming
a vector ${\vec \Delta}^M$ with components indexed by $L$, and
similarly the functions $\chi_{L_1L_2}^L(k)$ as composing a vector
${\vec\chi}^{L_1L_2}$.
The equations
\begin{equation}
\chi^{L_1L_2}_L(k)-\sum_{L'} (-1)^\Lambda \int
W_{LL'} (k,k')\frac{ \chi^{L_1L_2}_{L'}(k') } {2|\xi(k')|}{\rm d} \tau' =
\delta_{LL_1}\phi^{L_1L_2}(k)
\label{chiviaphi}
\end{equation}
for the shape factors $\chi^{L_1L_2}_L(k)$ are readily obtained
from eq.~(\ref{base1}).  From this result we see that
$\chi^{L_1L_2}_L(k_F) = \delta_{LL_1}$ for any $L_2$, since
the potentials $W_{LL'}(k,k')$ vanish for $k$ on the Fermi surface
while $\phi^{L_1L_2}(k_F)=1$ for any $L_1,L_2$. It is important to
observe that the equations (\ref{chiviaphi}) are free from any
log singularities and form a coupled system of {\it linear}
integral equations.  We further remark that these equations
can be solved {\it without regard} to the values of the coefficients
$D_{LL'}^{JM}=D_{LL'}^M$ given by the set of equations (\ref{defdf})
with $\Delta_{L_1}^{J_1M_1}(k)=\Delta_{L_1}^{M_1}$ expanded as
in (\ref{gen1}).

It is also important to take note of an essential difference in 
the reformulated equations (\ref{base1})--(\ref{chiviaphi}) as 
compared to the original gap equations (\ref{gengap}). The sum 
over orbital angular momenta $L_1$ in (\ref{gengap}) is in principle
infinite, while in (\ref{base1}) only the states with $L'=L$ and 
$L'=L\pm 2$ can contribute, due to the conservation of the total 
angular momentum $J$ and parity.  Therefore the number of components 
of the vectors ${\vec\Delta}^M$ and ${\vec\chi}^{L_1L_2}$ does not 
exceed 3, since $|L-L'|\leq 2$ and in eqs.~(\ref{base1})--(\ref{chiviaphi}) 
the total angular momentum $J$ has a fixed value.  We may infer that, 
in the case of a $^3$P$_2$--$^3$F$_2$ mixture where $J=2$, orbital 
angular momentum quantum numbers beyond 5 contribute only to the 
numerical factors $D^M_{L_1L_2}$.

It is instructive to write out eqs.~(\ref{base1}) and (\ref{chiviaphi})
in extended form for the $^3$P$_2$--$^3$F$_2$ problem.  Using a symbolic
notation in terms of the operator $T=W/ 2[\xi^2+\delta^2]^{1/2}$, the
system (11) reads
\begin{eqnarray}
\Delta^M_1+T_{11} \Delta^M_1+T_{13} \Delta^M_3&=&D^M_{11}\phi^{11}+D^M_{13}
\phi^{13} \, ,\nonumber \\
\Delta^M_3+T_{31} \Delta^M_1+T_{33} \Delta^M_3&=&D^M_{31}\phi^{31}+D^M_{33}
\phi^{33} \, .
\label{ds}
\end{eqnarray}
Suppose all the coefficients $D^M_{LL'}$ are zero except $D^M_{11}$; then the
solution ${\vec \Delta}^M$ of (\ref{ds}), a two-component vector,
reduces to
\begin{equation}
\Delta^M_L(k)=D^M_{11}\chi^{11}_L(k) \qquad (L=1,3)\, ,
\label{sol}
\end{equation}
where the components $\chi^{11}_1$ and $\chi^{11}_3$ of ${\vec \chi}_{11}$
obey the pair of equations
\begin{eqnarray}
\chi^{11}_1+T_{11} \chi^{11}_1+T_{13} \chi^{11}_3&=&\phi^{11}
\, , \nonumber \\
\chi^{11}_3+T_{31} \chi^{11}_1+T_{33} \chi^{11}_3&=&0 \, .
\end{eqnarray}
The vector ${\vec \chi}_{13}$ is given by a second pair of equations,
\begin{eqnarray}
\chi^{13}_1+T_{11} \chi^{13}_1+T_{13} \chi^{13}_3&=&\phi^{13}
\, , \nonumber \\
\chi^{13}_3+T_{31} \chi^{13}_1+T_{33} \chi^{13}_3&=&0 \, ,
\end{eqnarray}
derived by supposing that $D^M_{13}$ is present but all other
$D_{LL'}^M$ are zero.
The remaining two pairs of equations for the vector functions
${\vec\chi}^{31}$ and ${\vec\chi}^{33}$, namely
\begin{eqnarray}
\chi^{31}_1+T_{11} \chi^{31}_1+T_{13} \chi^{31}_3&=&0
\, , \nonumber \\
\chi^{31}_3+T_{31} \chi^{31}_1+T_{33} \chi^{31}_3&=&\phi^{31}
\, ,
\end{eqnarray}
and
\begin{eqnarray}
\chi^{33}_1+T_{11} \chi^{33}_1+T_{13} \chi^{33}_3&=&0
\, , \nonumber \\
\chi^{33}_3+T_{31} \chi^{33}_1+T_{33} \chi^{33}_3&=&\phi^{33}
\end{eqnarray}
are constructed in the same fashion.  The general solution of the 
$^3$P$_2$--$^3$F$_2$ pairing problem is given by eq.~(\ref{gen1}) 
with the indices $L_1$ and $L_2$ taking on the values 1 or 3.  It 
is worth noting that the same system of equations arises when one 
deals with $S-D$ pairing in the deuteron channel, except that the 
indices $L_1$ and $L_2$ then take on the values 0 and 2.

Let us return now to the problem of pairing in an arbitrary two-body
channel.  Having learned that the shape factors $\chi_L^{L_1L_2}(k)$ 
in the representation (\ref{gen1}) may be found by solving a set of 
nonsingular linear integral equations independently of the 
coefficients $D_{L_1L_2}^M$, we must next consider the determination 
of these coefficients, assuming knowledge of the $\chi$ functions.  
Substitution of (\ref{gen1}) into (\ref{defdf}) provides a set 
of nonlinear `algebraic' equations to be solved for the desired 
set of numbers:
\begin{eqnarray}
D^{JM}_{LL'}&=&(-1)^\Lambda v_{LL'}^J\int\phi^{LL'}(k)\sum_{L_1J_1M_1L_2L_3}
S_{L'L_1}^{JMJ_1M_1}({\bf n})D^{J_1M_1}_{L_2L_3} \nonumber \\
 &\qquad& \qquad \times \frac{\chi^{JL_2L_3}_{L_1}(k)}
{2\sqrt{\xi^2(k) + D^2({\bf k})}}{\rm d} {\bf n} {\rm d} \tau \, .
\label{deq1}
\end{eqnarray}
In the vicinity of the (upper) critical value $\rho_c$ of the density
$\rho$ where the gap value $\Delta_F$ vanishes, the term $D^2({\bf k})$
in the denominator of (\ref{deq1}) can be omitted, and in this case we
arrive at a system of {\it linear} equations for the coefficients
$D^{JM}_{LL'}$. The determinant of the linearized system should vanish at
$\rho=\rho_c$, since by definition there is no solution at that point.
Thus we obtain a new condition for gap closure.  For example, in the
case of $^3$P$_2$--$^3$F$_2$ pairing (where channel coupling is 
included), the zero of the determinant corresponding to the dominant 
$V_{11}^{J=2}$ matrix elements of the Argonne $v_{18}$ potential 
\cite{argonne18} gives a value $k_c = 3.6$ fm$^{-1}$ for the critical 
Fermi momentum when the input function $\xi(k)=\varepsilon(k)-\mu$ 
in eqs.~(\ref{deq1}) is constructed from free single-particle energies 
$\varepsilon(k)$.

In the standard numerical approach to BCS pairing problems, the original
system (\ref{gengap}) for the gap components $\Delta^{JM}_L(k)$
is solved by iteration.  One starts with suitable
``guesses'' for these functions (ten in number, for $^3$P$_2$--$^3$F$_2$
pairing) and inserts them into the r.h.s.\ of (\ref{gengap}), generating
a new set of component functions; and so on until convergence is
hopefully achieved.  Such a procedure encounters obstacles even in
the case of $^1$S$_0$ pairing \cite{kkc,baldo1,ccdk}.  A conventional
iterative approach is even more problematic for pairing in higher angular
momentum states, where the dimensionality of the nonlinear problem is
much larger.  From this point of view, Eqs.~(\ref{deq1}) are far more
appropriate as basis for numerical computation, since shape factors
$\chi_L^{L_1L_2}(k)$ are determined separately and iterations are needed 
only to find the set of coefficients $D^{JM}_{L_1L_2}$ entering the
representation (\ref{gen1}).

In principle, instead of trying to solve the system (\ref {deq1})
directly, we could use the variables $D_{LL'}^{JM}$ (twenty in number,
for $^3$P$_2$--$^3$F$_2$ pairing) to parametrize the condensation
energy and apply the variational principle to determine an appropriate
set of these factors and a corresponding set of gap components
$\Delta_L^{JM}(k)$.  However, one must remember that the performance of
optimization algorithms, as well as the convergence of iterative
procedures, is very sensitive to the degeneracies of the functional under
consideration.  Accordingly, any attempt at implementing such an
approach should be prefaced by a thorough study of the symmetries of
the gap problem and the degeneracies of its solutions.  Such a study is
of considerable fundamental interest in any case \cite{univ}.

\section{Triplet-P pairing without channel coupling\label{sec:nocoupling}}

To gain a better understanding of the triplet pairing problem in
neutron matter, we revisit the pure (i.e., uncoupled) $^3$P$_2$ 
case analyzed in some detail in \cite{tt71,ttr,ostgaard}.

If tensor coupling to the $^3$F$_2$ state is neglected, the system
of gap equations (\ref{gengap}) reduces to a set of five equations,
corresponding to $M = 0, \pm 1, \pm 2$. Taking into account
the property of time-reversal invariance, which implies the relations
\begin{equation}
\Bigl(\Delta^{JM}_L(k)\Bigr)^* = {(-1)}^{J+M} \Delta^{J,-M}_L(k) \, ,
\end{equation}
it is seen that only three of the generally complex components
$\Delta_1^{2M}$ are independent.  The independent components have
customarily been written as \cite{tt71,ttr,ostgaard}
\begin{eqnarray}
\Delta^{20}_1(k) &=& \delta_0(k) \, , \nonumber \\
\Delta^{21}_1(k) &=& \Bigl( \Delta^{2,-1}_1(k) \Bigr)^*
	= \delta_1(k) + i \eta_1(k) \, , \nonumber \\
\Delta^{22}_1(k) &=& \Bigl( \Delta^{2,-2}_1(k) \Bigr)^*
	= \delta_2(k) + i \eta_2(k) \, ,
\label{deltaeta}
\end{eqnarray}
where the $\delta_i(k)$ and $\eta_i(k)$ are real functions of $k$.

Takatsuka and Tamagaki \cite{tt71,ttr}, as well as Amundsen and
{\O}stgaard \cite {ostgaard}, have carried out numerical studies
of the pure $^3$P$_2$ problem that demonstrate an interesting
feature of triplet pairing in neutron matter (and in fact
of triplet-P pairing more generally).  Solutions of different types
were found to be nearly degenerate, i.e., they yield very nearly
the same condensation energy.  The spread of condensation energies
reported by Takatsuka and Tamagaki at a typical density is only 3\%,
and {\O}stgaard and Amundsen obtained very similar results.  The
calculations of these authors were based on an older nucleon-nucleon
potential called OPEG (see sec.~\ref{sec:tripletnumapp} for details).
Although this potential model is no longer competitive, its essential
aspects are semi-quantitatively correct.  At any rate,
it will be argued that the nearly degenerate character of the level
structure in the triplet pairing problem and the energy ordering
of different solution types are very insensitive to the interaction
assumed.

Important insights into the uncoupled problem can be achieved
within the separation method \cite {univ}.  For pure $^3$P$_2$ pairing,
eq.~(\ref{sol}) becomes simply
\begin{equation}
\Delta^{2M}_1(k)=D^{2M}_{11}\chi_1^{11}(k) \, ,
\label{dchi}
\end{equation}
while the shape factor $\chi_1^{11}(k)$ obeys the uncoupled equation
\begin{equation}
\chi_1^{11}(k) + \int \frac{W_{11}(k,k')\chi_1^{11}(k')}
        {2|\xi(k)|}{\rm d}\tau' =  \phi^{11}(k ) \,.
\label{samechi}
\end{equation}
We now make the decomposition 
$D^{M\neq 0}_{112}=(\lambda_M+i\kappa_M)\delta_0(k_F)/\sqrt{6}$,
where $\delta_0(k_F)=D^{M=0}_{112}$ serves as a scale
factor.  Relations between the parameters $\lambda_M$ and $\kappa_M$
and the quantities $\delta_i(k_F)$ and $\eta_i(k_F)$ ($i=1,2$) stem
from the definitions contained in eqs.~(\ref{deltaeta}).
Evaluating the spin sums $S^{2M2M'}_{11}$ and separating
real and imaginary parts in eq.~(\ref{deq1}), we extract the following
set of equations determining the coefficients $D_{11}^{JM}$
(cf.~refs.~\cite{univ,van}):
\begin{eqnarray}
    \lambda_2
&=& -v'[\lambda_2(J_0+J_5)-\lambda_1 J_1 -\kappa_1J_2-J_3]\, ,\nonumber \\
    \kappa_2
&=& -v'[\kappa_2(J_0+J_5)-\kappa_1 J_1 +\lambda_1J_2+J_4] \, ,\nonumber\\
    \lambda_1
&=& -v'[\lambda_1J_6-(\lambda_2+1)J_1+\kappa_2J_2-\kappa_1J_4/2]\, ,\nonumber\\
     \kappa_1
&=& -v'[\kappa_1J_7-\kappa_2J_1-(\lambda_2-1)J_2-\lambda_1J_4/2]\, ,\nonumber\\
1 &=&-v'[-(\lambda_1 J_1-\kappa_1 J_2+\lambda_2 J_3-\kappa_2 J_4)/3+J_5] \,,
\label{base2}
\end{eqnarray}
where $v'=(\pi/2)v_{11}^{J=2}$.  The integrals $J_i$ are given by
\cite{univ}
\begin{equation}
J_i = \int f_i(\theta,\varphi){\phi^{11}(k)\chi_1^{11}(k) \over
2E({\bf k})} k^2{\rm d} k {{\rm d }{\bf n}\over 4\pi} \, ,
\label{int10}
\end{equation}
with $f_0=1-3z^2$, $f_1=3xz/2$, $f_2=3yz/2$, $f_3=3(2x^2+z^2-1)/2$,
$f_4=3xy$, and $f_5=(1+3z^2)/2$, where $x=\sin\theta\cos\varphi$,
$y=\sin\theta\sin\varphi$, and $z=\cos\theta$.  Of these integrals,
only $J_5$ contains a principal term behaving like $\ln \Delta_F$, since
only $f_5$ among the $f_i$ does not vanish under the angular integration
when multiplied by an angle-independent factor.

The normal-state single-particle energy $\xi(k)$ present in the
energy denominator $E({\bf k})=[\xi^2(k)+D^2({\bf k})]^{1/2}$
is conventionally assumed to have finite slope $d\xi(k)/dk$, while
the gap term takes the explicit form
\begin{eqnarray}
D^2({\bf k})
&=& {{\delta_0^2}\over{16 \pi}}[(1+\lambda_2)^2+\kappa^2_1 +\kappa^2_2
     +(\lambda_1^2-4\lambda_2-\kappa^2_1)x^2 \nonumber\\
&-& 2(\lambda_1+ \lambda_1\lambda_2+\kappa_1\kappa_2) xz 
    +(3+\lambda^2_1-\lambda^2_2-2\lambda_2)z^2\nonumber\\
&+& 2(2\kappa_2-\kappa_1\lambda_1)xy
    +2(\kappa_1+ \lambda_1\kappa_2-\lambda_2\kappa_1)yz]
    [\chi_1^{11}(k)]^2 \, .
    \label{dsquare} 
\end{eqnarray}

In their angular content, the system (\ref{base2}) and the trigonometric
decomposition (\ref{dsquare}) are consistent with the formulations 
of the uncoupled $^3$P$_2$ problem given in refs.~\cite{ttr,tt71,ostgaard}, 
and specifically with eqs.~(4.7)--(4.8) of Amundsen and {\O}stgaard.
However, our separation formulation, as expressed in 
eqs.~(\ref{dchi})--(\ref{dsquare}),
has distinct advantages as a basis for numerical solution of this
problem.  To begin with, there is only {\it one} shape function
$\chi(k) \equiv\chi^{11}_1(k)$, which satisfies a linear integral
equation of exactly the same form as for singlet-S pairing
(cf.\ eq.~(\ref{chi}).  This property of shape invariance follows
naturally from the formal development presented in 
sect.~\ref{sec:sepformalism}.  Given the key physical fact that the
angle dependence of the problem only comes into play significantly in
the close vicinity of the Fermi surface, we observe that this part
of momentum space is almost entirely eliminated from the $k'$ integration
in eq.~(\ref{samechi}) by the vanishing of $W(k,k')$ at
$k'=k_F$.  Thus, to an excellent approximation, all solutions
$\Delta_1^{2M}$ of the uncoupled triplet problem have identical
shapes.  In particular, the positions of their zeros and maxima
coincide, as is clearly seen in the figures of previous works on
triplet pairing in neutron matter \cite{ttr,ostgaard}.

To establish the angular dependences of the allowed solutions of
the $^3$P$_2$ problem, the shape equation (\ref{samechi}) must of course
be supplemented by a set of equations that determine the coefficients or
amplitudes corresponding to the different $M$ values.  An important
benefit of the separation strategy is that the nonlinear aspects of
the problem are now concentrated entirely in the latter system of
equations, given by (\ref{base2}) --  equations for a set of
{\it numbers} $D_{112}^M$ rather than for a set of {\it functions}.
By contrast, earlier authors \cite{tt71,ttr,ostgaard} have worked directly
with a set of singular linear integral equations for the five functions
$\delta_0(k)$, $\delta_1(k)$, $\eta_1(k)$, $\delta_2(k)$, and $\eta_2(k)$
defined in eq.~(\ref{deltaeta}).  In this approach, the dependences of
gap quantities on the modulus of the momentum and on angles (or magnetic
quantum numbers) are confronted simultaneously, with attendant
opportunities for numerical difficulties.

Obviously, the process of numerical solution is facilitated by the
separation of ``shape'' and ``angular'' aspects of the problem that has 
been accomplished in eqs.~(\ref{dchi})--(\ref{dsquare}).  Less obviously,
this transformation opens the way to an incisive analysis of the nature
of these solutions.  Such an analysis has been carried out in
refs.~\cite{univ,van}, extending the formalism to finite temperature
$T$ via introduction of the factor $\tanh\left[E({\bf k})/k_B T\right]$
in the integrands of the basic gap equations eq.~(\ref{gengap}) and
the $J_i$ integrals (\ref{int10}).

Appealing to the orthogonality of the spherical functions for different
$M$, it is possible to find special solutions of the system (\ref{base2})
(or of the system (4.7) of ref.~\cite{ostgaard}) that satisfy most or
some of its equations identically.  The simplest types of consistent
solutions are characterized as follows \cite{tt71,ttr,ostgaard}:
\begin{enumerate}
\item[1.$~$] Only $M = \pm 2$ components are present:
$\delta_0=\delta_1=\eta_1=0$.
\item[2.$~$] Only $M = 0$ is present:
	$\delta_1=\eta_1=\delta_2=\eta_2=0$.
\item[3.$~$] Only $M = \pm 1$ are present:
	$\delta_0=\delta_2=\eta_2=0$.
\item[4.$~$] Coupled case containing $M = 0, \pm 2$:
	$\delta_1=\eta_1=0$.
\end{enumerate}
\noindent
In addition, there is naturally the general case involving all five 
$M$ values.

For purposes of illustration, let us examine the situation
for solution Types 1 and 2 in some detail.

{\it Type 1}.  
Under the assumption that only the component with $M=2$
is present (its $M=-2$ partner being included trivially), 
we need only deal with $\Delta_1^{22}(k)=\delta_2(k_F)\chi(k)$,
noting that a free choice of phase allows us to set $\eta_2=0$.
In this case only the first of eqs.~(\ref{base2}) is relevant and
it becomes
\begin{equation}
{1\over v} =- \int  \phi(k)\chi(k) {{3\over 2}\sin^2\theta\over
2\sqrt{\xi^2(k) +{3\over 2}\Delta_F^2\chi^2(k)\sin^2\theta}}
 {\rm d} {\bf n} {\rm d} \tau \, ,
\label{mtwo}
\end{equation}
where $v=v_{11}^{J=2}$.  All the other equations of the set 
(\ref{base2}) are met identically.  (We remind the reader that 
$\Delta_F^2$ is the value, on the Fermi surface, of the angle 
average defined in eq.~(\ref{aagap}).)  Apart from the angular
dependences and angular integration, eq.~(\ref{mtwo}) has the same form
as the equation (\ref{deltaf}) for the gap amplitude in the $^1$S$_0$
problem (see the appendix).  The angular integration can actually 
be performed analytically, with the result
\begin{equation}
{1\over v} =-  \int   {\phi(k)\chi(k) \over 2\sqrt{\xi^2(k) +
      \frac32\Delta^2_F\chi^2(k)}}\Phi_2(\beta_2){\rm d}\tau \, ,
\label{eqtype1}
\end{equation}
where
\begin{eqnarray}
\Phi_2(\beta_2) &=& \frac{3}{2\beta_2} \arcsin \beta_2 + \frac{3}{4\beta_2^2}
\left[\sqrt{1 - \beta_2^2}-  {\arcsin \beta_2\over\beta_2}\right] \, ,
\nonumber \\
(\beta_2(k))^2 &=& {{3\over 2}\Delta^2_F\chi^2(k)\over \xi^2(k) +
{3\over 2}\Delta_F^2\chi^2(k)}
\, .
\end{eqnarray}
The root of eq.~(\ref{eqtype1}) can be found numerically without difficulty,
after finding the universal shape function $\chi(k)=\chi_1^{11}(k)$
by matrix inversion of the discretized form of eq.~(\ref{samechi}).
This has been done for pairing matrix elements calculated from the 
OPEG potential employed by Takatsuka and Tamagaki \cite{tt71,ttr} and
Amundsen and {\O}stgaard \cite{ostgaard}; some results for this
case are reported in sect.~\ref{sec:tripletnumapp}.  

{\it Type 2}.  Here there is the single component
$\Delta_1^{20}(k)=\delta_0(k_F) \chi(k)= \Delta_F \chi(k)$.
The dispersion equation determining $\Delta_F$ is given by the last of
eqs.~(\ref{base2}), which becomes
\begin{equation}
{1\over v}=- \int \phi(k)\chi(k) {{1\over 2}(1 + 3 \cos^2\theta))\over
2\sqrt{\xi^2(k) + {1\over 2}\Delta_F^2(1 + 3\cos^2\theta)\chi^2(k)} }
{\rm d}{\bf n} {\rm d} \tau \, .
\label{mzero}
\end{equation}
Analytic integration over angles is again possible, yielding
\begin{equation}
 {1\over v} =-  \int  \frac{\phi(k)\chi(k) }{2\sqrt{\xi^2(k) +
   {1\over 4}   \Delta_F^2\chi^2(k)}}\Phi_0(\beta_0){\rm d}\tau \, ,
\label{eqtype2}
\end{equation}
where
\begin{eqnarray}
\Phi_0(\beta_0) &=&
\frac{1}{2\beta_0} \ln\left(\beta_0 + \sqrt{1 + \beta_0^2}\right)
 +  \frac{3}{4{\beta_0}^2}\left[\sqrt{1 + \beta_0^2}
- \ln\left(\beta_0 + \frac{1}{\beta_0}\sqrt{1 + \beta_0^2}\right)\right],
\nonumber \\
(\beta_0(k))^2 &=& {{3\over 4}\Delta_F^2\chi^2(k)\over
	\xi^2(k) + {1\over 4}\Delta_F^2\chi^2(k)} .
\end{eqnarray}
Some numerical results for this case are also reported in 
sect.~\ref{sec:tripletnumapp}, again for the OPEG potential.

\section{Small-parameter expansion and angle-average approximation
\label{sec:aaa}}

Quantitative description of pairing in nuclear systems is beset with 
special difficulties because the strong inner repulsion (or strong 
momentum dependence) present in some partial waves (notably S states)
implies that the momentum integral entering the gap equation
must be extended to large momenta \cite{baldo1,ccdk,kkc}.  This
behavior precludes the application of simple estimates such as
the BCS weak-coupling formula, which assumes sharp localization of the
pairing interaction around the Fermi surface.  However, important
simplifications are permitted by the existence of a small dimensionless
parameter $d_F$, given essentially by the scale of the energy gap
at the Fermi surface, relative to the value of the Fermi energy
$\varepsilon_F$.

In the case of $^1$S$_0$ pairing, the small parameter is just
$d_F = \Delta_F/\varepsilon_F$, where $\Delta_F\equiv \Delta(k_F)$.
Framing this problem within the separation method, Taylor-series
expansion of the energy denominator
of the equation for the gap amplitude $\Delta_F$ has
been employed in ref.~\cite{luso} to derive a sequence of approximants
for this quantity expressed as series of terms in
$(d_F^2)^n \ln d_F$ and $(d_F^2)^n$ with integral $n$.  The derivations 
involved are outlined in appendix A.  Even the leading (zeroth) 
approximant yields reasonable results: the exact solution for 
$\Delta_F$ within the relevant density range is reproduced within 
5\%, based on pairing matrix elements formed from the $^1$S$_0$ 
component of the Argonne $v_{18}$ interaction.

In the triplet case considered here, the small parameter is appropriately
defined as $d_F=[{\overline{D^2}}(k_F)]^{1/2}/ \varepsilon_F \equiv 
\Delta_F/\varepsilon_F$.  A characteristic value of this quantity in 
the density region of interest is 0.5 MeV/80 MeV $\simeq 6 \times 10^{-3}$, 
which is at least an order of magnitude smaller than typical values 
for singlet-S pairing.  On this basis, one may expect to achieve 
good accuracy through an approximation that retains only the 
principal logarithmic terms $\sim \ln D^2({\bf k})$ together with 
those independent of $D^2({\bf k})$, while omitting corrections 
$\sim d^2_F\ln d_F$. Small-parameter approximations are examined 
in more detail in the appendix.  Here we restrict ourselves only 
to brief remarks.

Consider the integral on the r.h.s.\ of eqs.~(\ref{deq1}) for the 
coefficients $D_{LL'}^{JM}$.  We may carry out an integration 
by parts with respect to the variable $\xi$, to derive a 
principal term (within the angular integration) that goes 
like $\ln D^2(k_F,{\bf n})$ and a remainder that involves 
$D^2({\bf k})$ only in the factor 
$$
\ln\left(|\xi| +\left[\xi^2+D^2({\bf k})\right]^{1/2}\right) \,.
$$
Since latter quantity is quite smooth in the vicinity of
the Fermi surface, its replacement by $\ln (2|\xi|)$ is valid
within corrections of order of $d_F^2\ln d_F$ that belong to
the next stage of approximation.  Writing $D^2(k_F,{\bf n})=
\Delta^2_F{\cal D}^2({\bf n})$, we extract the scale $\Delta_F^2$
and express the angular dependence by ${\cal D}^2({\bf n})$.
The dominant log term then decomposes as $\ln D^2(k_F,{\bf n})=
\ln \Delta^2_F+\ln {\cal D}^2({\bf n})$.  By virtue of the
orthogonality condition (\ref{ort1}), the angle-independent
portion $\ln \Delta^2_F$ contributes only to terms on the r.h.s.\
of eq.~(\ref{deq1}) that are diagonal in the magnetic quantum number,
i.e., only for $M_1=M$.  Since $\Delta_F$ is a small quantity, the
contribution from the scale term $\ln \Delta^2_F$ greatly exceeds that
from the angle-dependent part $\ln {\cal D}^2({\bf n})$, which
remains of order of 1 even in the limit $\Delta_F\rightarrow 0$.   It
is worth noting that the $\Delta$-independent contributions coming
from terms in $\ln (2|\xi|)$ are also of order of 1.  However, these
terms are angle-independent and, like the scale term, contribute
only to the diagonal part of the r.h.s.\ of (\ref{deq1}).  In fact,
to within corrections of order $\sim d^2_F\ln d_F$, the result for
the $D$ coefficients calculated from the angle-independent terms
alone (i.e., dropping $\ln {\cal D}^2({\bf n})$), is the {\it same} 
as the result we obtain if we simply replace the quantity $D^2({\bf k})$ 
in the energy denominator of the original equation by its angle 
average (\ref{aagap}), 
\begin{equation}
{\overline{D^2}}(k) =
\frac1{8\pi} \sum_L |\Delta^{JM=0}_L(k)|^2 ~ +
\frac1{4\pi} \sum_{LM>0} |\Delta^{JM}_L(k)|^2 \, .
\label{avgap}
\end{equation}

We reiterate that justification of this replacement, which corresponds
to the {\it angle-average approximation} first introduced by Baldo
et al.~\cite{baldo2}, rests on (i) the unimportance of the angle-dependent
quantity $\ln {\cal D}^2 ({\bf n})$ relative to $\ln \Delta_F^2$ and
(ii) the unimportance of correction terms of order $d_F^2 \ln d_F$
or higher in the small parameter expansion.  The angle-average
approximation generally overestimates the pairing gap and the magnitude
of the pairing energy (i.e.\ the pair condensation energy correction to
the energy of the normal ground state).  However, it allows a considerable
simplification of the triplet pairing problem with very modest
sacrifice in accuracy, if our primary interest lies in the
magnitude of the pairing effect rather than the detailed structure
of pairing solutions.

Making the replacement (\ref{avgap}) in the general coupled-channel
equations (\ref{deq1}) for the gap amplitudes, we arrive at the
quasi-uniform system
\begin{equation}
D_{LL'}^M =  (-1)^\Lambda v_{LL'}\int\phi^{LL'}(k)\sum_{L_1L_2}
D_{L_1L_2}^M \frac{\chi^{L_1L_2}_{L'}(k)}
{2\sqrt{\xi^2(k) + {\overline{D^2}}(k)}}{\rm d}\tau \, .
\label{deq2}
\end{equation}
Since the index $M$ appears explicitly only on the quantities
being sought, there exists a degeneracy with respect to this
quantum number:  All sets of $D_{LL'}^{JM}$ yielding the same
result for the gap quantity ${\overline{D^2}}(k)$ of eq.~(\ref{avgap})
are equivalent.

\begin{figure}
\begin{center}
\psfig{figure=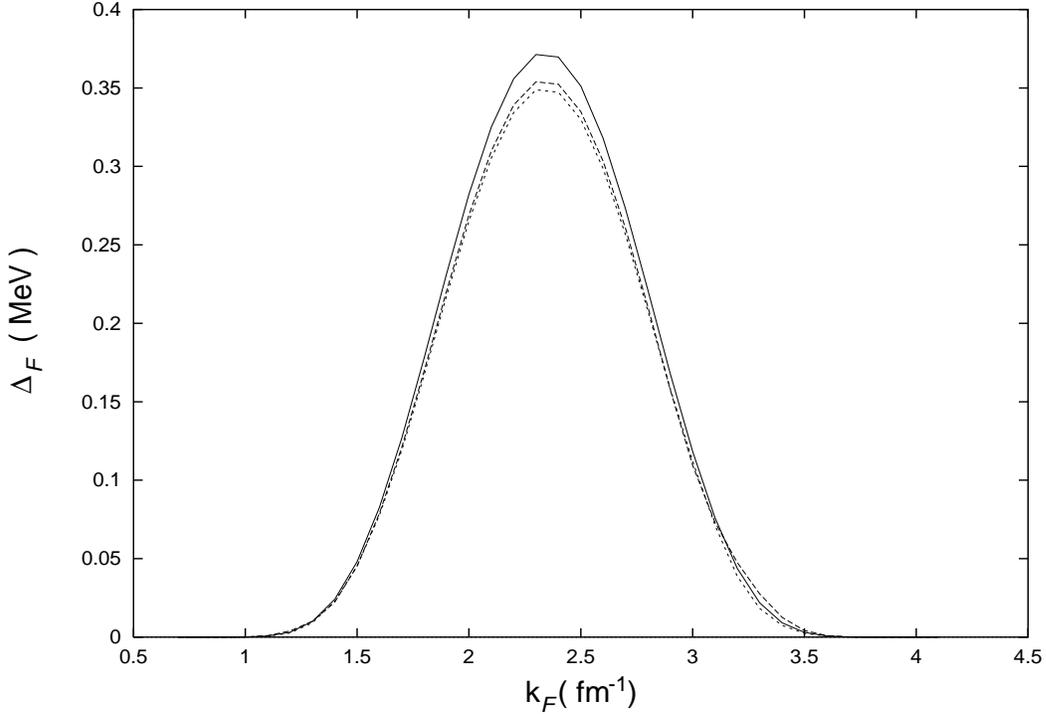,height=3.7truein,width=5.5truein,angle=-90}
\end{center}
\caption{
Test of the angle-average approximation for $^3$P$_2$ pairing, without 
coupling to the $^3$F$_2$ state.  The energy gap parameter $\Delta_F$ 
(square root of the quantity (\ref{aagap}), evaluated at the Fermi surface) 
is plotted versus Fermi momentum $k_F$, for the pure $M=0$ solution 
(long-dashed curve) and the $|M|=2$ solution (short-dashed curve), and 
as calculated in the angle-average approximation (solid curve).  The pairing 
matrix elements are constructed from the OPEG potential, and 
free single-particle energies are assumed ($m^*=1$).
}
\label{fig:aaaandm}
\end{figure}

We first apply the angle-averaging prescription to the pure
$^3$P$_2$ problem.  In this case eq.~(\ref{deq2}) becomes simply
\begin{equation}
1=-v  \int  { \left(\phi^{11}(k)\right)^2\chi^{11}_1(k) \over
   2\sqrt{\xi^2(k) + {\overline{D^2}}(k)}}{\rm d}\tau
\label{sss}
\end{equation}
with ${\overline{D^2}}(k) =\Delta^2_F[\chi^{11}_1(k)]^2 
=\Delta^2_F\chi^2(k)$ and $v=v_{11}^{J=2}$.
This equation, having no explicit reference to magnetic substates,
is identical in form to the singlet-channel equation (\ref{deltaf})
and is solved by the same method.  Numerical results from solution
of the uncoupled gap problem in this approximation are presented
and discussed in sect.~\ref{sec:tripletnumapp}.
If we do not care about the detailed angular behavior of solutions, but 
only about the size of the gap (as measured by $\Delta_F$) and the
condensation energy, the angular-averaging prescription gives very
good predictions.  This statement is justified by the close
agreement with the results shown in fig.~\ref{fig:aaaandm} for
solution Types 1 and 2 and by the near degeneracy of the
$\Delta_F$ values and condensation energies found numerically
\cite{tt71,ttr,ostgaard} and analytically \cite{univ} for the
different solution types.

Turning to the general coupled-channel case, observe that the
terms with $L_1 \neq L'$ on the r.h.s.\ of (\ref{deq2}) contain
integrals
\begin{equation}
A_{LL'}^{L_1L_2}=\int \frac {\phi^{LL'}(k)\chi^{L_1L_2}_{L'}(k)}
{2\sqrt{\xi^2(k) + {\overline{D^2}}(k)}} {\rm d} \tau
\approx \int \frac{\phi^{LL'}(k)\chi_{L'}^{L_1L_2}(k)}{2|\xi(k)|}
{\rm d} \tau
\end{equation}
which are almost independent of the gap value, since the property
$\chi_{L'}^{L_1L_2}=\delta_{LL_1}$ implies that the numerators of
the integrands vanish at the Fermi surface. In the diagonal case,
$L_1=L'$, we introduce an unknown positive number
\begin{equation}
Z = \int \frac{[\chi^{11}_1(k)]^2}
{2\sqrt{\xi^2(k) + {\overline{D^2}}(k)}} {\rm d} \tau \label{zdef}
\end{equation}
and write
\begin{equation}
\int \frac{\phi^{LL'}(k)\chi^{L'L_2}_{L'}(k)}{2\sqrt{\xi^2(k)
+ {\overline{D^2}}(k)}} {\rm d} \tau \equiv
Z + B^{L'L_2}_{LL'} \, ,  \label{bdef}
\end{equation}
where
\begin{equation}
B^{L'L_2}_{LL'}=
\int \frac {\phi^{LL'}(k)\chi^{L'L_2}_{L'}(k) -[\chi^{11}_1(k)]^2 }
{2\sqrt{\xi^2(k) + {\overline{D^2}}(k)}} {\rm d} \tau \approx
\int \frac
{\phi^{LL'}(k)\chi^{L'L_2}_{L'}(k) -[\chi^{11}_1(k)]^2 }
{2|\xi(k)|} {\rm d} \tau \, .
\label{bform}
\end{equation}
The simplification on the right in eq.~(\ref{bform}) is permitted
since $\phi^{LL'}(k_F)= \chi^{L'L_2}_{L'}(k_F)=1$ and the
numerator in the integrands again vanishes at the Fermi surface.
The integrals $B^{L'L_2}_{LL'}$ are effectively already known:
with the indicated simplifications, they only depend on the
$\chi$ functions, which may be found from eqs.~(\ref{chiviaphi})
without reference to the gap quantities $D_{LL'}^M$ being sought.

We now specialize to coupled-channel, $^3$P$_2$--$^3$F$_2$ pairing 
where the system (\ref{deq2}) for the coefficients $D_{LL'}^M$
contains four equations.  For a solution of this set of equations 
to exist, the determinant
\begin{equation}
C(Z) =
\left|
\begin{array}{cccc}
(1/v_{11}) + B^{11}_{11} + Z & B^{13}_{11} + Z & A^{31}_{11}
& A^{33}_{11} \\
A^{11}_{13} & -(1/v_{13}) + A^{13}_{13} & B^{31}_{13} + Z
& B^{33}_{13} + Z \\
B^{11}_{31} + Z & B^{13}_{31} + Z & -(1/v_{31}) + A^{31}_{31}
& A^{33}_{31} \\
A^{11}_{33} & A^{31}_{33} & B^{31}_{33} + Z & (1/v_{33}) +
B^{33}_{33} + Z
\end{array}
\right| \, ,\label{det}
\end{equation}
which depends on $\Delta_F$ (or the $D_{LL'}^M$ quantities) only
through the parameter $Z$, must vanish.
Evaluation of this determinant may be simplified by subtracting the
third row from the second row and the second row from
the fourth row, and then subtracting the second column
from the first column and the third column from the fourth column.
This leads to a {\it quadratic} equation for the determination
of the unknown parameter $Z$:
\begin{equation}
C(Z) = c_2 Z^2 + c_1 Z + c_0 = 0 \, ,
\label{viet}
\end{equation}
where
\begin{equation}
c_0 = C(0) =
\left|
\begin{array}{cccc}
C_{00} & C_{01} & C_{02} & C_{03} \\
C_{10} & C_{11} & C_{12} & C_{13} \\
C_{20} & C_{21} & C_{22} & C_{23} \\
C_{30} & C_{31} & C_{32} & C_{33}
\end{array}
\right| \, ,
\end{equation}
\begin{equation}
c_2 =
\left|
\begin{array}{cc}
C_{30} - C_{10} - [C_{31} - C_{11}] & C_{33} - C_{13} - [C_{32} - C_{12}]\\
C_{00} - C_{20} - [C_{01} - C_{21}] & C_{03} - C_{23} - [C_{02} - C_{22}]
\end{array}
\right| \, ,
\end{equation}
\begin{equation}
c_1 = C(1) - c_0 - c_2 \, .
\end{equation}
Equation (\ref{viet}) can have two roots $\{Z^\pm_0\}$, and
only positive solutions are of interest.

Having found the appropriate root of (\ref{viet}), one may solve
any three of the four equations of the system (\ref{deq2})
for the ratios $D_{LL'}^M/D_{11}^M$.  These ratios may then be
used to obtain an equation for $D_{11}^M$ that is analogous to
the algebraic equation for the gap amplitude $\Delta_F$ in
the $^1$S$_0$ problem, namely
\begin{equation}
Z = \int
\frac
{{[\chi^{11}_1(k)]}^2}
{2\sqrt{\xi^2(k) + {\overline{D^2}}(k)}} {\rm d}\tau\, .
\end{equation}
Solution of this equation yields the desired gap function in
angle-average approximation.

The angle-average approximation serves well if the goal is to calculate
the gap amplitude $\Delta_F = [{\overline{D^2}}(k_F)]^{1/2}$.  However,
if information is required on the spectrum of pairing solutions
and the energy splittings between these solutions, it is essential
to include the angle-dependent effects from the terms in eq.~(\ref{deq1})
going like $\ln{\cal D}^2({\bf n}) $.  By so doing, accurate
results and valuable insights can be gained even if we stop at
leading order in the small-parameter approximation.

The situation is made more tangible by considering two simple
solutions involving only one $M$ value (and its negative), specifically
$M=2$ or $M=0$ (corresponding to Types 1 and 2, respectively).
Working at lowest order in the small-parameter expansion,
partial integration of the definition (\ref{int10}) of the integrals
$J_i$ over the variable $\xi$ leads to the simplified
form
\begin{equation}
J_i=-{1\over 2}\int_0^1\int_0^{2\pi}f_i(z,\varphi)\ln D^2(z,\varphi)
{{\rm d}z{\rm d} \varphi\over 2\pi} \, ,
\label{jexp}
\end{equation}
having made use of the equalities $\phi(\xi=0)=\chi(\xi=0)=1$.
With this result, eqs.~(\ref{mtwo}) and (\ref{mzero}) are recast,
respectively, as
\begin{equation}
1=-vJ_5\left(\Delta_F^{(M=2)}\right) \quad {\rm and}
\quad 1=-v\left[J_0\left(\Delta_F^{(M=0)}\right)
+J_5\left(\Delta_F^{(M=0)}\right)\right] \, .
\end{equation}
The splitting between the states in question is determined by subtracting
one equation from the other.  Straightforward manipulations give
\begin{equation}
\ln\left({\Delta_F^{(2)}\over \Delta_F^{(0)}}\right)^2
={2\over 3}-\ln 3+{2\pi\over 9\sqrt{3}}\simeq -0.028 \, .
\label{ratio}
\end{equation}

When the angle-average approximation is applied, eq.~(\ref{mtwo})
or (\ref{mzero}) is replaced by
\begin{equation}
\frac{8\pi^2\epsilon_F}{v k^3_F} 
	+ \int_{1}^\infty {\rm d} \epsilon
	\frac{\sqrt{\epsilon+1}\phi\chi}{2\epsilon}
	+ \int^{1}_{-1}{\rm d}\epsilon
		\frac{\sqrt{\epsilon+1}\phi\chi-1}{2|\epsilon|}
	+ \ln \frac4{\Delta^2_F} - \ln \frac1{8\pi} = 0 \, ,
\label{aaa0}
\end{equation}
which is to be solved for $\Delta_F = \Delta_F^{\rm aa}$.
(Here, $\phi=\phi(\epsilon)$ and $\chi=\chi(\epsilon)$ after
a change of integration variable from $k$ to $\epsilon=(k/k_F)^2-1$.)
Evaluating the ratios of $\Delta_F^{(2)}$ and $\Delta_F^{(1)}$
to $\Delta_F^{\rm aa}$ along the lines of the above derivation of
$\Delta_F^{(2)}/\Delta_F^{(0)}$, the ordering of the three
gap amplitudes is seen to be $\Delta_F^{(2)} < \Delta_F^{(0)}
< \Delta_F^{\rm aa}$.  This order agrees with the numerical results 
appearing in fig.~\ref{fig:aaaandm} and with the findings of other
authors \cite{tt71,ttr,ostgaard}.

It is interesting and highly significant that the result (\ref{ratio}) 
for the ratio of pairing energies for the Type 1 ($|M|=0$) and Type 2 
($M=0$) solutions -- obtained to leading order in the small parameter 
of the theory yet very accurate---is completely independent of the 
parameters of the actual interaction and independent of density.
In other words, it is {\it universal}.

Proceeding to the next layer of complexity, let us consider the
Type 4 solution, again studied in leading order.  In this case,
$M=0$ and $|M|=2$ substates coexist, and the system of equations
that determines the coefficients $D_{11}^{JM}$ becomes
\begin{equation}
\lambda_2=-v\left[\lambda_2(J_0+J_5)-J_3\right] \, , \label{20lam}
\end{equation}
\begin{equation}
1=-v\left[-\lambda_2J_3/3+J_5\right] \, ,
\label{scale}
\end{equation}
where we have introduced the more compact notation
$\lambda_2=\delta_2\sqrt{6}/\delta_0$ and the $J_i$ are given
by (\ref{int10}). The first of these equations can be conveniently
rewritten as
\begin{equation}
0=\lambda_2J_0+(\lambda^2_2/3-1)J_3
\label{dif}
\end{equation}
by subtracting $\lambda_2$ times eq.~(\ref{scale}) from
eq.~(\ref{20lam}).

The alternative form (\ref{dif}) is distinctly advantageous: unlike
$J_5$, the integrals $J_0$ and $J_3$ do not contain a term
proportional to $\ln\Delta_F^2$ and hence are independent of
any details of interaction.  Consequently, eq.~(\ref{dif}) provides
a closed equation for the parameter $\lambda_2$ whose solution
is universal -- i.e., the same for all pairing interactions.
Once the universal parameter $\lambda_2$ is determined, the scale
parameter $\delta_0$ is calculated from (\ref{scale}); its value does,
of course, depend on the interaction.

The formulas for $J_0$ and $J_3$ can be further simplified by partial
integration of $J_0$ over $z$ and $J_3$ over $\varphi$.  The remaining
integration over $\varphi$ is performed analytically.  After simple
operations, we arrive at
\begin{equation}
J_0(0,\lambda_2)=-{1\over 3}+ 4\int_0^1{z^2{\rm d}z \over f(z)} \, ,
\quad
J_3(0,\lambda_2)={3+\lambda_2^2\over 4\lambda_2}-
{3\over 8\lambda_2}\int_0^1 f(z){\rm d}z\, ,
\end{equation}
where
$f(z)=\left[(1+\lambda^2_2+(3-\lambda^2_2)z^2)^2
-24\lambda_2^2(1-z^2)^2\right]^{1/2}$.
In general, these integrals are expressible in elliptic functions, but
in exceptional cases they reduce to elementary integrals.
This is exactly what happens when we seek roots of the
system (\ref{20lam}--(\ref{dif}).  In addition to
the root $\lambda_2=0$ corresponding to the $M=0$ state (the
Type 2 solution), we find the roots $\lambda_2=\pm 3$ and
$\lambda_2=\pm 1$.  It should not be surprising that the
roots are degenerate with respect to signs, since eq.~(\ref{dif})
is invariant under reversal of the sign of $\lambda_2$.
What is most surprising is that these roots are rational numbers.

Rationality of the roots $\lambda_2$ is a consequence of a hidden
symmetry of the problem that transcends the approximation we have
employed.  To substantiate this assertion, consider first the particular
solution $\lambda_2=3$, for which the squared gap function
(\ref{dsquare}) becomes
$D^2(k,x,z;\lambda_2=3)\sim [\chi_1^{11}(k)]^2[4-3(x^2+z^2)]$.
The symmetry of this function with respect to $x$ and $z$ implies
the relation $3J_0(0,3)+2J_3(0,3)=0$, since the combination
$3f_0+2f_3=6(x^2-z^2)$ changes sign on interchange
of $x$ and $z$ whereas $d^2$ and other factors within the integrand of
(\ref{int10}) are left unchanged. The above relation between $J_0$ and $J_3$
satisfies eq.~(\ref{dif}).  The essential point is that this relation
derives from a simple symmetry of the problem; it is {\it not} 
predicated on any approximate scheme for the evaluation of the $J_i$.
Now consider the root $\lambda_1=0$, for which
$D^2(k,x,y,z)\sim [\chi_1^{11}(k)]^2(1-x^2)$. Since 
$D^2(k,x,y,z)$ is independent of $y$ and $z$, one may trivially 
integrate (\ref{int10}) over these variables.  Simple algebra then 
gives $3J_0(0,1)=2J_3(0,1)$, and this relation also satisfies 
Eq.~(\ref{dif}).  We may therefore conclude that the ratio of 
$M=2$ and $M=0$ components in the Type 4 solution is independent 
of any input parameters including density, temperature, and the 
parameters specifying the pairing interaction.

A general analysis along similar lines, again exploiting
the simplifications produced by the separation method,
reveals the fundamental structure of the full set of solutions
of the BCS $^3$P$_2$ pairing problem in neutron matter \cite{univ}.
Considering general multicomponent solutions,
the allowed ratios of coefficients of different spin-angle
matrices ${\hat G}^{JM}_L({\bf  n})$ in the expansion (\ref{matgap})
of the gap matrix turn out to be universal.  As in the
above derivation, any reference to interaction, density, or
temperature cancels out to very high accuracy in forming the ratios.
The complete set of solutions for these ratios is obtained
explicitly by solving a set of five coupled `algebraic' equations,
one of which serves to determine the energetic scale of the
pairing effect.  The solution set and the spectrum of pairing
energies are found to be highly degenerate in character.  The
small splittings between different solutions can be found by
manipulations similar to those performed above for the Type 1
and Type 2 solutions.

In principle, analogous analytic arguments may be used to establish
universalities of pairing in the much more difficult problem of
liquid $^3$He, where states with $L=S=1$ and $J=0,1,2$ contribute
on an equal footing and the number of equations to be solved
rises from five to nine.

\section{Numerical applications: uncoupled and coupled channels
\label{sec:tripletnumapp}}

The separation approach to BCS-type pairing problems which has been 
further developed and studied in this paper leads to considerable 
simplifications of the process of solving the gap equations in a generic 
pairing channel.  Aided by the existence of a small parameter, 
this approach allows one to find the {\it interval of existence} of 
solutions without having to perform any iterations,
and it allows one actually to find {\it solutions} without having to 
iterate a set of integral equations.  Some iterations may be required
to construct solutions, but they need be performed only on a set of 
algebraic equations for the amplitudes of the components of the gap 
matrix (quantities that are momentum-independent but dependent on
the magnetic projection $M$).  The ``division of labor'' that underlies 
the separation method thus facilitates the process of finding solutions;
moreover, because of the reduced number of numerical operations, 
it permits higher accuracy to be achieved.  In addition to these numerical 
advantages, the new method offers an efficient framework for fundamental 
analytic investigations of pairing, as has been demonstrated in 
sect.~\ref{sec:nocoupling} and more systematically in 
refs.~\cite{kkc,univ,luso,van}.

A necessary step in testing our procedures for numerical solution 
of the triplet pairing problem in neutron matter is comparison with 
results obtained by other authors.  As indicated in
sect.~\ref{sec:nocoupling}, Takatsuka and Tamagaki \cite{tt71}
(see also their review \cite{ttr}) have studied the pure $^3$P$_2$ 
pairing problem at some depth, classifying the types of solutions
and generating numerical solutions by iteration.  Subsequently,
Takatsuka \cite{t72} (see also \cite{ttr}) implemented an iterative 
procedure for solution of the gap equations in the $^3$P$_2$--$^3$F$_2$ 
coupled-channel case with fixed $M = 0$.  This early numerical work 
was based on several older potential models \cite{opeg} with soft 
(Gaussian) cores.  The Takatsuka-Tamagaki calculations were repeated 
and supplemented by Amundsen and {\O}stgaard \cite{ostgaard}. 
Baldo et~al.~\cite{baldo2} introduced the idea of angular averaging in 
considering $M=0$ and $|M|=2$ solutions (i.e., Type 2 and Type 1 
solutions) of the pure $^3$P$_2$ case and reported preliminary numerical 
results for a separable approximation to the Argonne $v_{14}$ potential 
\cite{argonne14}.  More recently, Elgar{\o}y et~al.~\cite{elgaroy} 
have performed uncoupled $^3$P$_2$ and coupled $^3$P$_2$--$^3$F$_2$ 
calculations with fixed $|M|$ (0 or 2) in $\beta-$stable matter for the 
Bonn potential models designated A, B, and C \cite{bonnold1,bonnold2}.  

In the most recent microscopic study of triplet pairing, Baldo 
et al.~\cite{catoslo} obtained results for several realistic models 
of the nucleon-nucleon interaction, including four that may legitimately 
be regarded as phase-shift equivalent.  The latter potentials -- Argonne 
$v_{18}$ \cite{argonne18}, CD-Bonn \cite{bonnnew}, and Nijmegen I and 
II \cite{nijmegen} -- belong to a new generation of models that 
yield high-precision fits of the sanitized world supply of $pp$ and 
$np$ scattering data up to a bombarding energy of 350 MeV.  All four 
produce fits with $\chi^2$ per data point close to unity.  In addition, 
gap calculations were performed for three ``older'' potentials -- 
Argonne $v_{14}$, Paris \cite{paris}, and Bonn B. 
The authors of this extensive study conclude that reliable predictions 
of the $^3$P$_2$--$^3$F$_2$ pairing gap at densities above about
1.7 times the saturation density of symmetrical nuclear matter
will require potential models that fit the nucleon-nucleon phase 
shifts up to 1 GeV.

For the purpose of comparison with the pre-1990 results on the triplet 
pairing problem, we have chosen the OPEG (``one-pion-exchange Gaussian'') 
potential \cite{opeg} used in refs.~\cite{tt71,t72,ttr,ostgaard}, and 
specifically the version denoted OPEG $^3O-1$.  This potential has been 
fitted to scattering data collected before 1970, and it cannot be 
regarded as a realistic model of the nucleon-nucleon interaction.  
However, it possesses all of the essential qualitative features of 
the interaction and describes them at a semi-quantitative level.  Of 
local character, it contains central, tensor, spin-orbit 
(${\bf L} \cdot {\bf S}$), and quadratic spin-orbit 
(e.g.~$({\bf L} \cdot {\bf S})^2$) components, as written
out in detail by Amundsen and {\O}stgaard \cite{ostgaard}.

\begin{figure}
\begin{center}
\psfig{figure=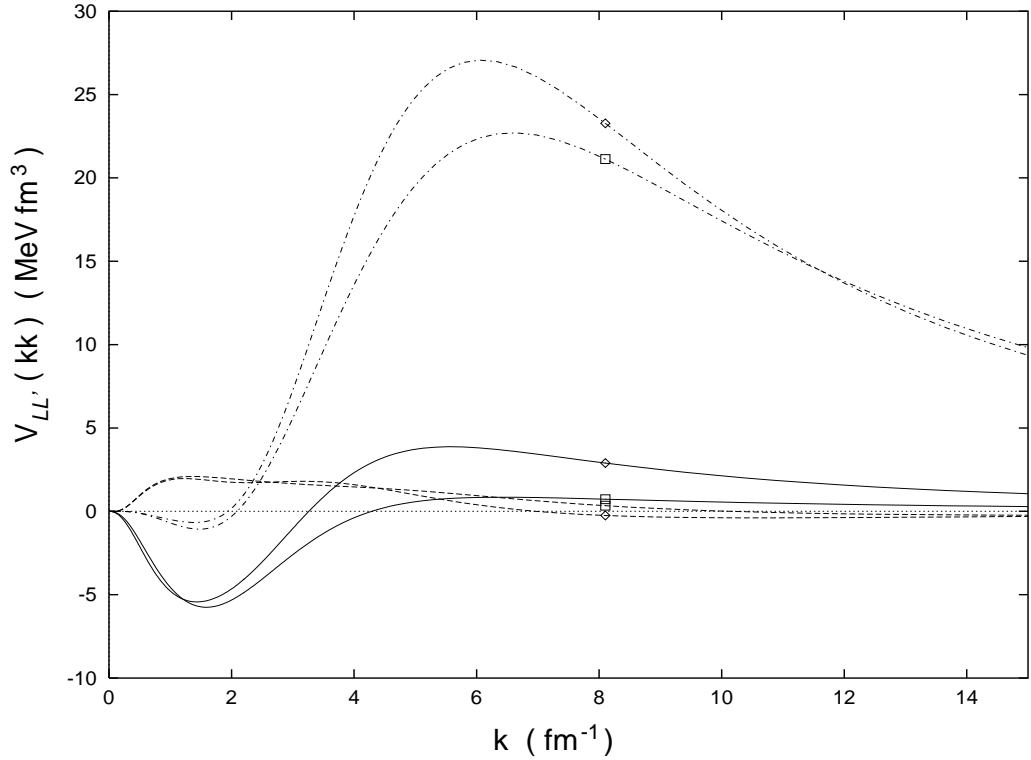,height=3.7truein,width=5.5truein,angle=-90}
\end{center}
\caption{
Diagonal pairing matrix elements $V_{LL'}^{J=2}(k,k)=V_{11}(kk)$ (solid
curves), $V_{13}(kk)$ (dashed curves), and $V_{33}(kk)$
(dot-dashed curves) in the $^3$P$_2$--$^3$F$_2$ coupled channel, as 
functions of momentum $k$.  Curves for the Argonne $v_{14}$ and $v_{18}$
interactions are labeled with a square and with a diamond, respectively.
}
\label{fig:diagme}
\end{figure}

To compare with the latest calculations \cite{catoslo}, we have 
generated results for the Argonne $v_{14}$ and Argonne $v_{18}$ models.
The diagonal pairing matrix elements of these two potentials 
are plotted in fig.~\ref{fig:diagme}.  

In presenting the numerical results, we employ two measures of
the gap matrix, namely (i) the average quantity $\Delta_F \equiv 
\left[{\overline{D^2}}(k_F)\right]^{1/2}$ defined via eq.~(\ref{aagap})
or eq.~(\ref{avgap}) and (ii) the gap component $\Delta_L^{JM}(k)$, 
either as a function of the momentum modulus $k$ or evaluated at the 
Fermi momentum $k_F$.  We note that when the pairing problem is solved 
in the angle-average approximation specified in sect.~\ref{sec:aaa}, 
the $M$ dependence of the latter measure becomes moot.  Thus the
index $M$ will often be suppressed, along with the fixed index
$J=2$.

\begin{figure}
\begin{center}
\psfig{figure=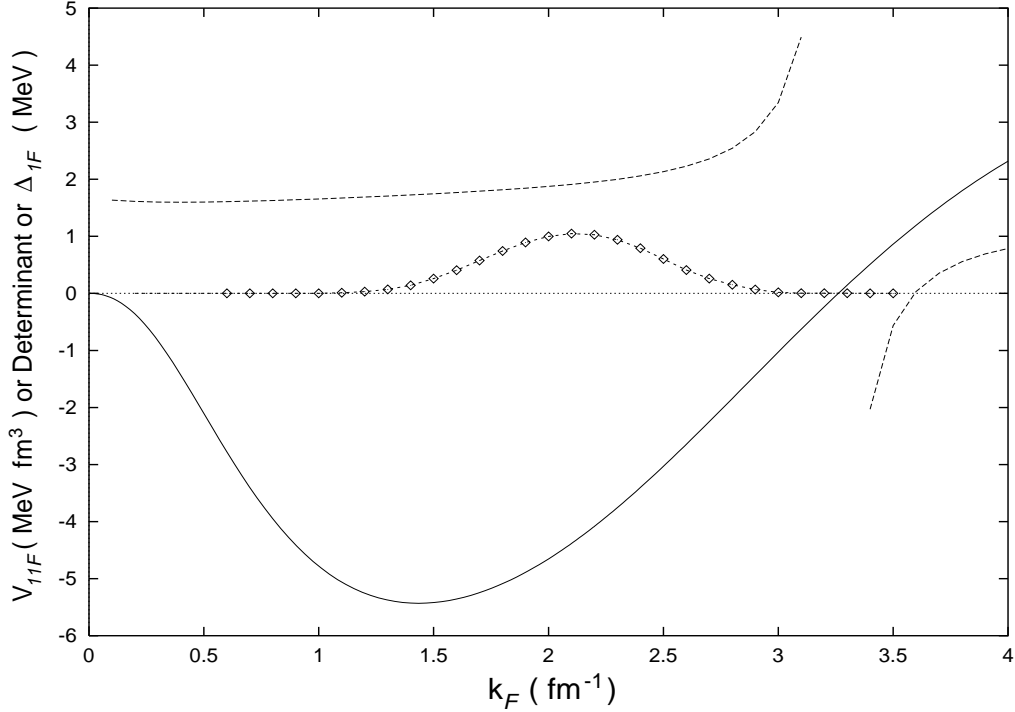,height=3.7truein,width=5.5truein,angle=-90}
\end{center}
\caption{
Illustration of gap existence and closure for $^3$P$_2$ pairing
by the Argonne $v_{18}$ interaction, with coupling to the $^3$P$_2$ 
state turned off.  Dependences on Fermi momentum $k_F$, of the diagonal 
matrix elements $V_{11}^{J=2}(k_F,k_F)\equiv V_{11F}$ of the pairing 
interaction (solid curve), the dimensionless characteristic determinant 
of the linearized version of system (\ref{deq1}) (dashed curve), and 
the gap component value $\Delta_1^{2M}(k_F)\equiv\Delta_{1F}$ 
determined in angle-average approximation (dotted curve 
with diamonds).  Free single-particle energies are assumed.
}
\label{fig:3in1}
\end{figure}

The zeros of the characteristic determinant of the linearized version
of eq.~(\ref{deq1}) provide us with candidate values for the upper 
critical Fermi momentum $k_c$.  This is true whether we restrict 
ourselves to pure $^3$P$_2$ pairing or address the coupled-channel
$^3$P$_2$--$^3$F$_2$ case.   Examining first the uncoupled case, 
in fig.~\ref{fig:3in1} we superimpose plots of the diagonal matrix
elements $v_{11}^{J=2}= V_{11}^{J=2}(k_F,k_F)$ of the Argonne $v_{18}$
potential, the associated characteristic determinant, and the gap 
component $\Delta_1^{2M}(k_F)=\Delta_1(k_F)$ obtained in the 
angle-average approximation, all considered as functions of the Fermi 
momentum $k_F$.  From the preceding formal developments it is 
evident that a zero of the potential implies a pole of $\phi^{11}(k)$ 
and hence a simple pole of the determinant.  Such a pole is very prominent 
in fig.~\ref{fig:3in1} and is naturally followed closely by a zero of the 
determinant.  The position of this zero gives a value 
$k_c \simeq 3.6$ fm$^{-1}$ for the upper critical $k_F$ value at 
which the gap closes, the corresponding upper critical density being 
$\rho_c= k_c^3/3 \pi^2$.  On the plot, the gap amplitude 
is clearly negligible at this point.  Because of the exponential 
falloff, it appears to be zero even below this point, but is in 
fact still finite over an interval in $k_F$ where $v^{J=2}_{11}$ has 
positive values.

\begin{figure}
\begin{center}
\psfig{figure=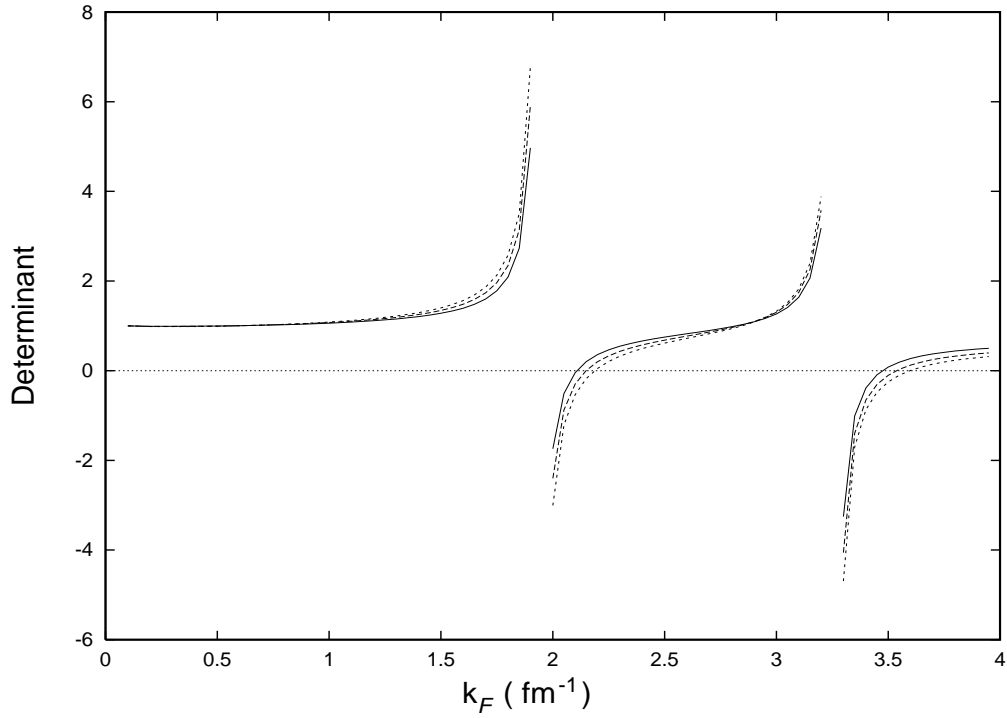,height=3.7truein,width=5.5truein,angle=-90}
\end{center}
\caption{
The characteristic determinant for pairing in the $^3$P$_2$--$^3$F$_2$ 
coupled channel as a function of Fermi momentum $k_F$.  Results are
shown for the Argonne $v_{18}$ potential and three values of the parameter 
$m^*=M_n^*/M_n$ in an effective-mass approximation for the 
normal-state single-particle energies: $m^*=1$ (short-dashed curve), 
$m^*=0.8$ (long-dashed curve), and $m^*=0.6$ (solid curve).}
\label{fig:tripdet}
\end{figure}

Fig.~\ref{fig:tripdet} displays the characteristic determinant in 
the coupled-channel case, again with the Argonne $v_{18}$ potential 
as input for the pairing force.  The determinant is evaluated for 
three different effective-mass ratios $m^*=M_n^*/M_n$, the free 
normal-state single-particle spectrum $\varepsilon(k)=\hbar^2 k^2/2M_n$ 
being replaced by $\hbar^2 k^2/2M_n^*$ to simulate dispersive effects of 
the medium.  For any of the chosen values of $m^*$, the determinant 
has a pole near $k_F=1.9$ fm$^{-1}$ followed closely by a zero near 
$k_F=2.2$ fm$^{-1}$, and a second pole near $k_F=3.3$ fm$^{-1}$ 
followed closely by a zero near $k_F=3.5$ fm$^{-1}$.  It is the 
second zero that should be identified with the upper critical density,
since the adjacent pole corresponds to a zero of $v_{11}^{J=2}$.
The point is that we are seeking to describe $T=1$
triplet-P-wave pairing as modified by the tensor coupling to the
$T=1$ triplet-F channel.  The first pole of the determinant
corresponds to the zero of $v_{33}^{J=2}$.  Thus the accompanying
determinantal zero is not relevant to the task at hand, which
is the identification of the upper critical point for triplet-P
rather than triplet-F pairing.  Generally, a {\it lower bound} 
on $k_c$ for pairing predominantly in a state with orbital angular 
momentum quantum number $L$ is given by the location of the pole 
that is produced by the vanishing of $v^J_{LL}$ (we assume that
this occurs only at one density).   Conventional wisdom gives
$v^J_{LL}<0$ as a sufficient (but not necessary!) condition
for a pairing instability in partial wave $L$.  The indicated lower 
bound property is of course consistent with this statement, but we 
are able to provide a more incisive condition for the extent of the 
pairing instability in terms of the behavior of the characteristic 
determinant associated with the integral equation (\ref{deq1}) 
(or alternatively, with the characteristic determinant of the 
integral equation (\ref{chiviaphi}) for the shape function 
$\chi(k)$ \cite{luso,vvkthesis}).  Concerning the dependence
of $k_c$ on effective mass, it is seen in fig.~\ref{fig:tripdet}
that a smaller $m^*$ leads to a lower value of the upper critical 
Fermi momentum (cf.\ further discussion below).

\begin{figure}
\begin{center}
\psfig{figure=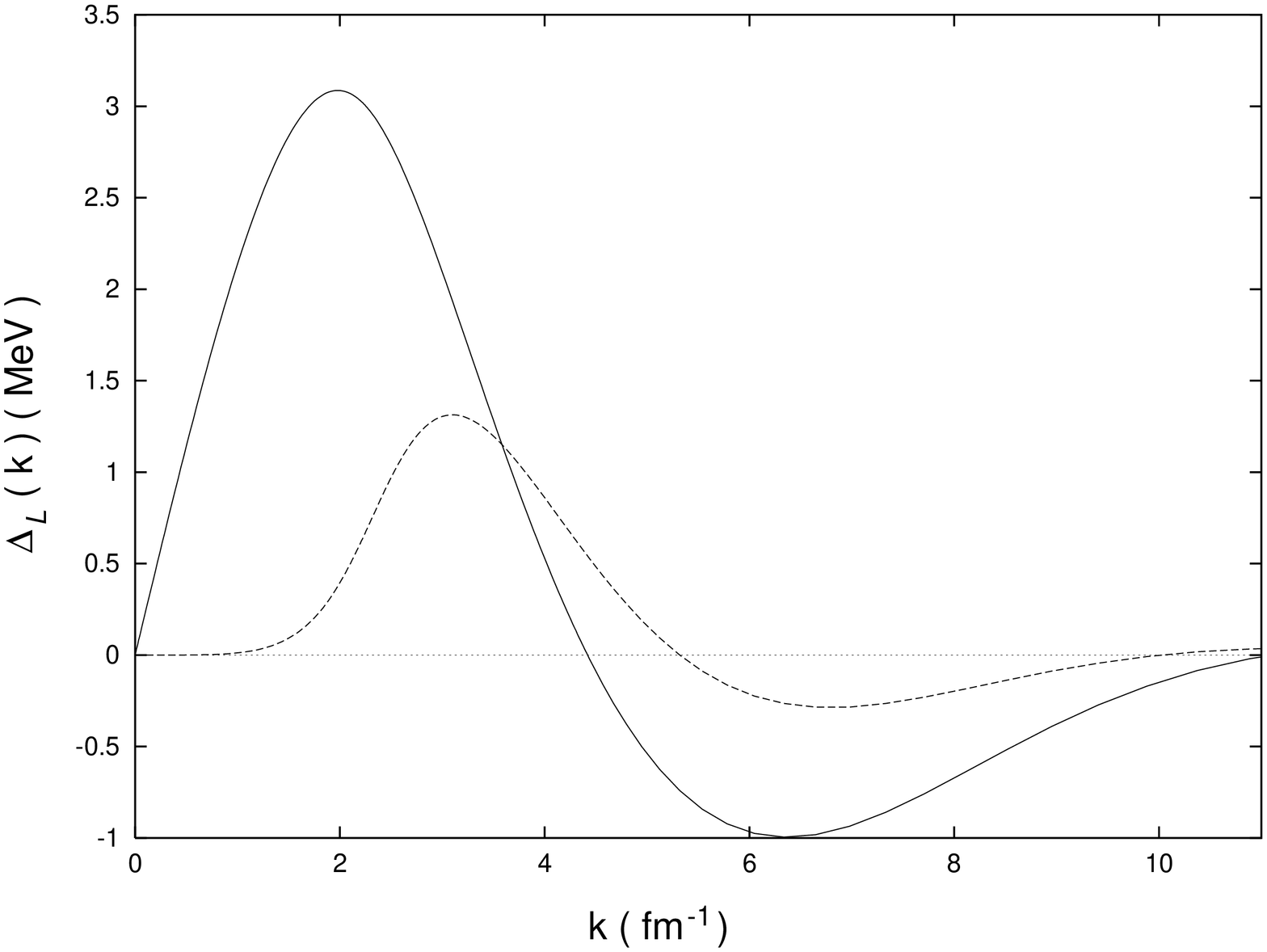,height=3.7truein,width=5.5truein,angle=-90}
\end{center}
\caption{
Gap components $\Delta_1^{2M}(k)\equiv\Delta_1(k)$ (solid curve) and 
$\Delta_3^{2M}(k)\equiv\Delta_3(k)$ (dashed curve) for 
$^3$P$_2$--$^3$F$_2$ pairing (coupled-channel case),
calculated in angle-average approximation for the OPEG potential
and $m^*=1$, and plotted versus momentum $k$ at $k_F = 2.2$ fm$^{-1}$. 
}
\label{fig:opegcoupled}
\end{figure}

Numerical results for the OPEG interaction that may be compared with 
those of Takatsuka and Tamagaki and Amundsen and {\O}stgaard (and with 
general findings of other authors) are displayed in fig.~\ref{fig:aaaandm}
and fig.~\ref{fig:opegcoupled}.  Fig.~\ref{fig:aaaandm} refers to 
the uncoupled $^3$P$_2$ case with $m^*=1$ and shows results for the 
$k_F$ dependence of values of the energy-gap parameter $\Delta_F$ 
corresponding, respectively, to the angle-average approximation (uppermost 
curve), the pure $M=0$ solution (middle curve), and the $|M|=2$ solution 
(lowest curve).  The latter two curves are in reasonable agreement with 
results reported in refs.~\cite{ttr,tt71,ostgaard}, having maxima near 
2 MeV in the vicinity of $k_F = 2.2$ fm$^{-1}$.  It is worth mentioning 
again that our calculations are in accord with the energy ordering of the 
$M=0$ and $|M|=2$ solutions found numerically in the early work 
\cite{tt71,ttr,ostgaard}; moreover, they confirm 
the generic results that have been derived in sect.~\ref{sec:aaa} 
using the separation method and the small parameter expansion.  At 
$k_F=2.2$ fm$^{-1}$ the ratio predicted by eq.~(\ref{ratio}) is 
reproduced with better than 1\% accuracy, and the energy order 
$\Delta_F^{(2)} < \Delta_F^{(0)} < \Delta_F^{\rm aa}$ implied by 
eqs.~(\ref{ratio}) and (\ref{aaa0}) is preserved.  Considering the 
near-degeneracies in the condensation energies for the different 
solutions, documented originally by Takatsuka and Tamagaki \cite{tt71}, 
these favorable comparisons warrant some confidence in the application 
of the angle-average approximation to the coupled-channel case.

Introducing the coupling to the $^3$F$_2$ channel, we have solved the
triplet pairing problem for the OPEG and Argonne potentials in the 
angle-average approximation, using the procedure outlined in 
sect.~\ref{sec:aaa}.  Fig.~\ref{fig:opegcoupled} shows the 
calculated momentum dependence of the gap components 
$\Delta_1^{2M}(k)\equiv\Delta_1(k)$ (solid curve) and 
$\Delta_3^{2M}(k)\equiv\Delta_3(k)$ (dashed curve) for the case
of the OPEG potential and $m^*=1$, at a Fermi momentum $k_F=2.2$ fm$^{-1}$
near the value yielding the maximum strength of triplet pairing.  The 
locations of the maxima, minima, and nodes of these functions are in close 
agreement with results given in refs.~\cite{t72,ttr}, and the magnitudes 
are in general agreement.  (It is to be noted that in his coupled-channel 
calculation, Takatsuka included only the $M=0$ magnetic substate.)

\begin{figure}
\begin{center}
\psfig{figure=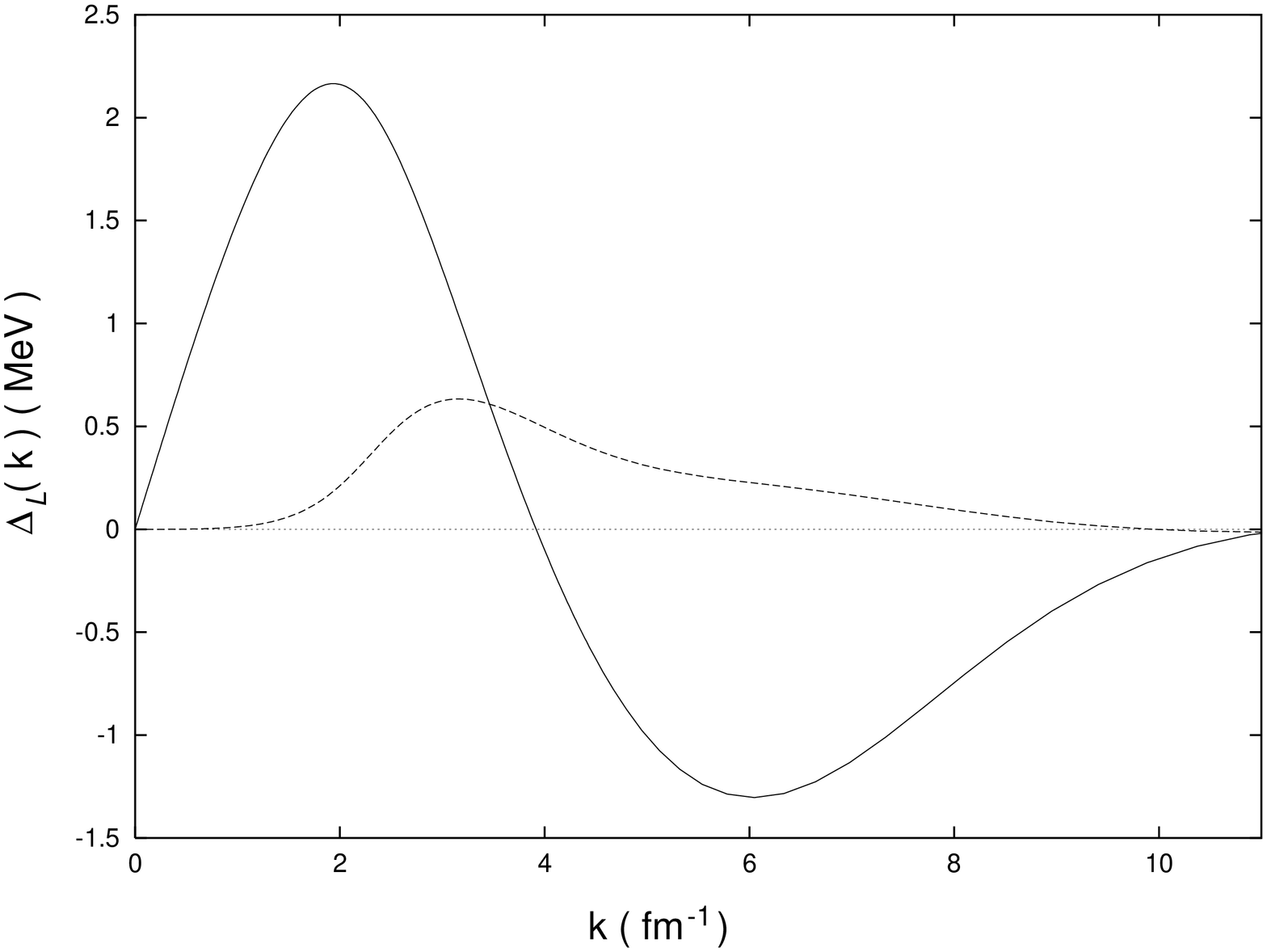,height=3.7truein,width=5.5truein,angle=-90}
\end{center}
\caption{
Gap components $\Delta_1^{2M}(k)\equiv \Delta_1(k)$ (solid curve) and 
$\Delta_3^{2M}(k)\equiv \Delta_3(k)$ (dashed curve) for $^3$P$_2$--$^3$F$_2$ 
pairing (coupled-channel case), calculated in angle-average approximation 
for the Argonne $v_{18}$ potential and $m^*=1$, and plotted versus momentum 
$k$ at $k_F = 2.2$ fm$^{-1}$. 
}
\label{fig:v18coupled}
\end{figure}

\begin{figure}
\begin{center}
\psfig{figure=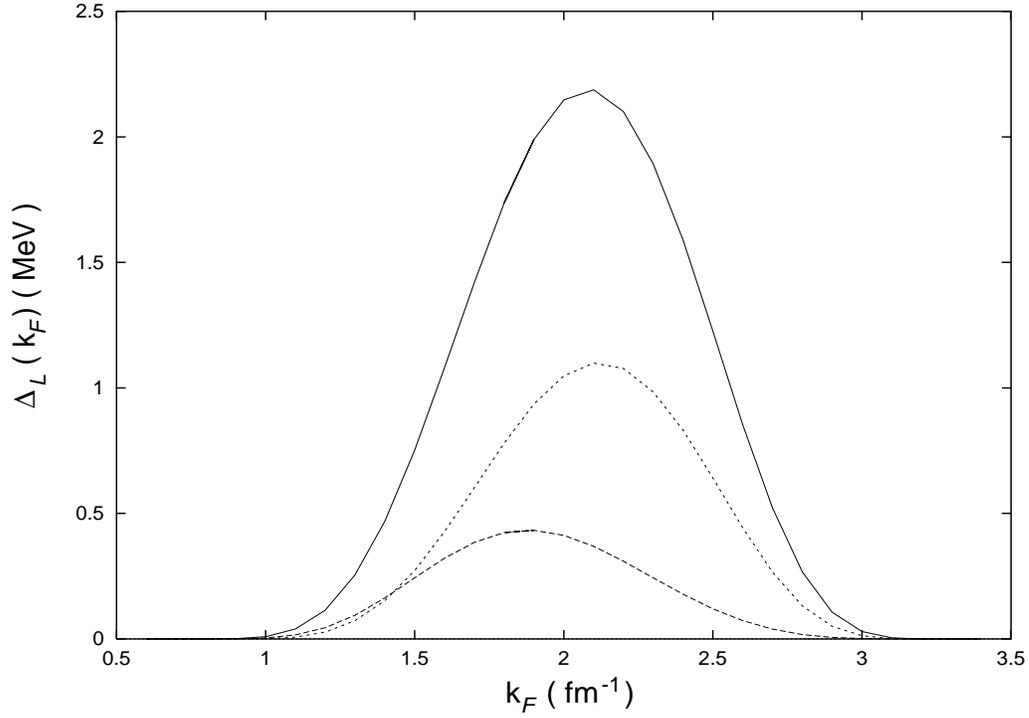,height=3.7truein,width=5.5truein,angle=-90}
\end{center}
\caption{
Gap-component values $\Delta_1^{2M}(k_F)\equiv \Delta_1(k_F)$ (solid curve) 
and $\Delta_3^{2M}(k_F)\equiv \Delta_3(k_F)$ (long-dashed curve) {\it with} 
channel coupling, and $\Delta_1(k_F)$ {\it without} channel coupling 
(short-dashed curve), calculated in angle-average approximation for the 
Argonne $v_{18}$ potential and $m^*=1$, and plotted as a function of Fermi 
momentum $k_F$.
}
\label{fig:deltav18}
\end{figure}

\begin{figure}
\begin{center}
\psfig{figure=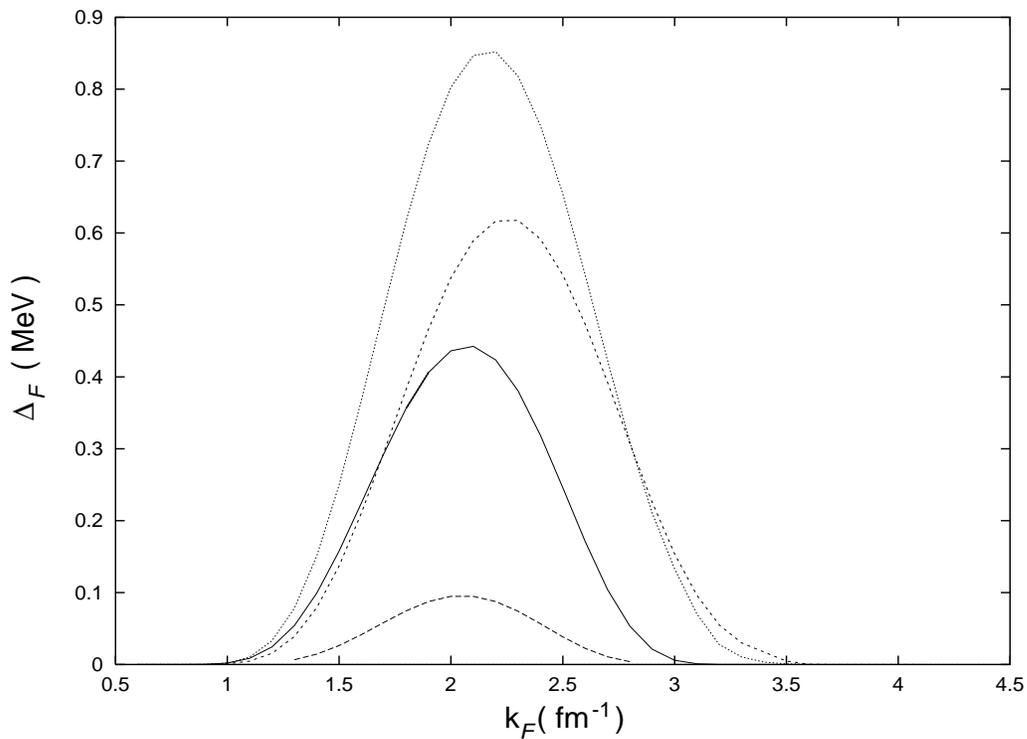,height=3.7truein,width=5.5truein,angle=-90}
\end{center}
\caption{
Dependence on Fermi momentum $k_F$ of the overall gap measure 
$\Delta_F$ (square root of the quantity (\ref{aagap}), evaluated 
at the Fermi surface) for coupled-channel $^3$P$_2$--$^3$F$_2$ 
neutron pairing, as determined in the angle-average approximation.  
Dotted curve: for Argonne $v_{14}$ interaction and effective-mass 
parameter $m^*=1$.  Short-dashed curve:  for OPEG potential and 
$m^*=1$.  Solid curve: for Argonne $v_{18}$ potential and 
$m^*=1$.  Long-dashed curve: for Argonne $v_{18}$ interaction and 
$m^*=0.78$.  
}
\label{fig:opegandv18coupled}
\end{figure}

Figs.~\ref{fig:v18coupled}, \ref{fig:deltav18}, and 
\ref{fig:opegandv18coupled} depict the
principal results of our coupled-channel $^3$P$_2$--$^3$F$_2$ 
gap calculations for the more realistic potential models,
$v_{18}$ and $v_{14}$.  Fig.~\ref{fig:v18coupled} illustrates the momentum 
dependence of the $L=1$ and $L=3$ components $\Delta_L^{2M}(k)$ for the
Argonne $v_{18}$ interaction (abbreviated as $\Delta_1(k)$ and
$\Delta_2(k)$).  As expected, the shapes of these gap 
functions are rather similar to those found for the OPEG interaction
at the same density (cf.~fig.~\ref{fig:opegcoupled}).  However, the
overall scale of the pairing effect in the OPEG case is almost
50\% larger than for the $v_{18}$ model.

In fig.~\ref{fig:deltav18}, we view the dependence of the gap 
component values $\Delta_2(k_F)$ and $\Delta_3(k_F)$ 
on the Fermi momentum $k_F$, as calculated in angle-average approximation 
for the $v_{18}$ interaction.  Contrasting the results for 
$\Delta_1(k_F)$ with and without coupling to the $F$ wave, 
it is seen that the tensor force acts to substantially increase the 
pairing effect, as has been demonstrated in greater or lesser
measure by earlier investigators \cite{t72,ttr,ostgaard,elgaroy}.

Fig.~\ref{fig:opegandv18coupled} compares the $k_F$ dependence
of the $^3$P$_2$--$^3$F$_2$ energy-gap parameter $\Delta_F$ for 
the three potentials investigated, the angle-average approximation 
being invoked in all cases.  In all three cases, the gap measure 
$\Delta_F$ peaks in the range between $k_F=2$ and 2.5 fm$^{-1}$.  With
the free single-particle spectrum, i.e., $m^*=1$, the maximum value 
reached by this quantity for the $v_{14}$ interaction is twice 
that attained for the more refined $v_{18}$ model.  As is clear
from its definition, $\Delta_F$ contains direct contributions both from 
P-wave and F-wave pairing.  Inspecting fig.~\ref{fig:diagme}, we
observe that the greater pairing effect for the $v_{14}$ potential 
relative to $v_{18}$ can be traced to the greater attraction present 
for $v_{14}$ in both P- and F-wave $J=2$ channels over the relevant 
range of the momentum variable $k$.  (The coupling matrix elements
$V_{13}(kk)$ for the two potentials are essentially equivalent 
over the latter range.)  Fig.~\ref{fig:opegandv18coupled} also 
demonstrates the greater pairing effect of the OPEG potential 
relative to $v_{18}$.   Such discrepancies between the predictions
obtained for different interaction models are not unexpected:  due  
to the log singularity intrinsic to the BCS pairing phenomenon,
the overall strength $\Delta_F$ of the effect in the $^3$P$_2$--$^3$F$_2$
channel is extremely sensitive to the behavior of the input pairing 
matrix elements at momenta in the vicinity of the Fermi surface.  In 
turn, since the pairing matrix elements are constructed directly
from the given free-space $NN$ potential model and since 
$k_F \sim (2-3)$ fm$^{-1}$, one must expect a sensitive dependence on 
properties of the bare interaction that are not determined, or 
only poorly determined, by fits of the $NN$ scattering data 
up to 350 MeV.  This point has been developed thoroughly and 
convincingly in the recent work of Baldo et~al.~\cite{catoslo}.

In table 1, we report some explicit numerical values of $\Delta_F$
obtained in our calculations based on the two ``realistic'' potential 
models, Argonne $v_{14}$ and $v_{18}$.  A comparison of our numerical 
results for these potentials with the corresponding results of Baldo 
et~al.~\cite{catoslo} is in order, restricting attention perforce to 
the case of free single-particle energies $\varepsilon(k)$.   First, we 
point out that the pairing matrix elements entering our formulas are larger 
than those appearing in ref.~\cite{catoslo} by a numerical factor 
$\pi/2$, due to different choices of normalization for the spherical 
Bessel functions.  Upon accounting for this (immaterial) factor, the 
diagonal matrix elements plotted for the $v_{18}$ interaction in fig.~3 
of ref.~\cite{catoslo} are apparently in excellent agreement with 
their counterparts in fig.~\ref{fig:diagme}.  Second, it must be
recognized that the quantity $D^2(k)$ defined in eq.~(6) of 
ref.~\cite{catoslo} is actually twice the function $D^2(k)$
appearing in our treatment (see our eqs.~(\ref{squd}) and 
(\ref{avgap})).  Thus the gap measure $\Delta_F\equiv 
[{\overline{D^2}}(k_F)]^{1/2}$ we have calculated is
smaller than the $\Delta_F$ quantity of Baldo et al.\ by a factor 
$1/\sqrt{2}$.  The definition we have adopted is motivated
by the desideratum of consistency with the earlier literature 
\cite{ttr,ostgaard}.  In any case, the two calculations should 
agree once this scale factor is applied, since both calculations 
employ the angle-average approximation.

The results are generally in good agreement; however, there
are some modest but significant numerical discrepancies.
For the $v_{14}$ interaction, the current results for
$\Delta_F$ versus $k_F$ are somewhat higher than those of 
ref.~\cite{catoslo} below the peak, and somewhat lower above.  
The position of the peak, occurring near $k_F=2.3$ fm$^{-1}$ in 
the calculation of Baldo et al., is shifted to a slightly smaller 
$k_F$ value, near 2.2 fm$^{-1}$.  The value of $\Delta_F$ at the peak is 
higher in our calculation by roughly 5\%.  In the case of the 
$v_{18}$ interaction, the numerical agreement is excellent below 
the peak, but beyond the maximum the current results lie below those 
of ref.~\cite{catoslo} by an amount that can reach about 0.1 MeV 
(in the $\Delta_F$ measure used by Baldo et~al.).  

The origin of these relatively minor numerical disparities
is not clear.  We would, however, like to reiterate the concern
expressed in refs.~\cite{elgaroy,catoslo} about the sensitivity
of the numerical integration over $k$ in the triplet gap equations
to the choice of mesh points near the Fermi momentum.  The
difficulty arises because the functions $k^2 \Delta_1^{2M}(k)/E(k)$
and $k^2 \Delta_3^{2M}(k)/E(k)$ are very strongly peaked at
$k_F$.  To handle this problem within the present treatment,
we have integrated out the troublesome log-like part by hand
and performed a numerical integration only for the remainder,
which is much smoother and well-behaved.  For the latter integration,
we employ 288 points, dividing the $k$ domain into three intervals:
$[0,k_F]$, $[k_F,2k_F]$, and $[k_F,k_{\rm max}]$, with 
$k_{\rm max}= 50$ fm$^{-1}$.  Each interval contains 96 mesh
points distributed in Gaussian fashion -- very dense around the
ends of the interval and sparse in the middle.

In fig.~\ref{fig:opegandv18coupled} and table 1, we have included 
$v_{18}$ results not only for the baseline choice $m^*=1$, but also 
for a value $m^*=0.78$ of the effective-mass parameter conventionally
thought \cite{t72,ttr} to provide a sensible representation
of the in-medium single-particle energy.  This brings us to
the last point that we would like to address in the current
treatment of triplet pairing in neutron matter.  
Relative to the bare-mass case, the propagator renormalization implied 
by $m^*=0.78$ yields a substantially smaller gap as well as a lower 
value for the upper critical density (see fig.~\ref{fig:tripdet}).
These effects can be readily understood in terms of a reduction of
the density of states $N(0)$ near the Fermi surface, which is related 
to the effective mass by the familiar expression
$N(0)= k_FM_n^* / \pi^2 \hbar^2$.  In the singlet-S-wave problem, the
approximate formula (\ref{df0}) derived in the appendix shows clearly 
the inverse exponential dependence of the leading component of 
the gap on the density of states in the case of a negative pairing 
interaction on the Fermi surface, as does the BCS weak-coupling 
formula in a more limited context.  The same physical reasoning and 
similar mathematical connections apply in the triplet case.  

It is a common belief that at the higher densities where
triplet pairing becomes important, the interactions with particles in 
the background neutron sea produce a considerably greater reduction 
from the bare mass than is the case at inner-crust densities where 
singlet pairing dominates \cite{ttr,chao}.  This conclusion rests on 
the results of the Brueckner-Hartree-Fock and correlated-basis 
calculations. However, in these calculations, the contribution of 
pion-exchange effects to the renormalization of the nucleon effective 
mass is ignored.  The role of the pion degrees of freedom grows rapidly 
with increasing density, eventually giving rise to a rearrangement 
of the ground state (associated, in particular, with pion condensation 
\cite{migdalpi,pireview}).  The latest variational prediction \cite{akmal} 
of the density $\rho_c$ at which this phase transition sets in is based 
on a nuclear Hamiltonian containing the Argonne $v_{18}$ two-nucleon 
interaction (with boost corrections to account for the leading relativistic 
effects) and phenomenological three-nucleon interactions.  The result 
$\rho_c\sim 0.2$ fm$^{-3}$ actually lies below the density at which 
triplet pairing is expected to reach its maximum strength.  Straightforward 
considerations \cite{dyugaev1,dyugaev2} demonstrate that in the vicinity 
of this transition point the contribution of the pion-exchange 
diagram to the effective mass $M_n^*$ diverges and $m^*$ goes to 
infinity.  Therefore it is not unlikely that the operative value 
of the effective mass in the domain of triplet pairing considerably 
exceeds the bare mass.  From this standpoint, conventional treatments
that assume an $m^*$ value somewhat below unity may lose their relevance.

\section{Prospectus \label{sec:concl}}
Exploiting the separation approach and the existence of a small
parameter, we have been able to clarify important general properties
of triplet pairing in neuron matter and to establish a framework and
explicit procedures for reliable numerical solution of the relevant 
BCS gap equations.  This effort complements a recent exhaustive analysis of 
the full set of generic solutions of the pure (uncoupled) $^3$P$_2$ problem, 
also performed with the aid of the separation strategy \cite{univ}.  In view 
of the corresponding advances that have been made in the $^1$S$_0$ pairing 
problem \cite{kkc,luso,jensen}, it can be reasonably claimed that the 
{\it mathematical and analytical} aspects of BCS pairing of identical 
fermions in $^1$S$_0$ and $^3$P$_2$ states are now well understood, both 
within the relevant nucleonic contexts and in a more universal, 
context-independent sense.  The same cannot be said of the specific 
{\it physical and mechanistic} aspects of triplet or singlet pairing 
of strongly-interacting fermions, which are associated with the realistic 
in-medium particle-particle interaction and renormalized single-particle 
energies that should provide the raw material for the gap equations.  
A complete and realistic treatment of pairing in a given strongly-coupled 
Fermi system demands {\it ab initio} calculation of these essential
inputs.  

The success of the method of correlated-basis (or CBF) theory across 
a broad range of many-body systems and phenomena \cite{bishop,ccdk} 
suggests the following approach to the quantitative physics of pairing in
extended nucleonic systems:
\begin{enumerate}
\item[(a)]
Dressing of the pairing interaction by Jastrow correlations within 
CBF theory \cite{kroclark,ksj}
\item[(b)]
Dressing of the pairing interaction by dynamical collective effects
within CBF theory \cite{ksj,chen} (including polarization effects 
arising from exchange of density and spin-density fluctuations, etc.) 
\item[(c)]
Consistent renormalization of single-particle energies by short-
and long-range correlations within CBF theory (cf.\ ref.~\cite{kcj})
\end{enumerate}
This approach has already been explored in the $^1$S$_0$ neutron pairing
problem \cite{chen,ccdk}, although the assumed Jastrow correlations
have not been optimized and only a second-order CBF perturbation
treatment is available for step (b).  Application of this scheme 
to $^3$P$_2$--$^3$F$_2$ pairing in neutron-star matter is 
a challenging but potentially rewarding prospect.  

The inclusion of other physical effects will also be important in
achieving a realistic description of pairing in the neutron-star
medium.  Modifications due to three-nucleon interactions should
be considered, the calculations should be performed in $\beta$-stable
matter (cf.\ ref.~\cite{elgaroy}), and relativistic effects should 
be examined (see, for example, ref.~\cite{elgaroyrel1,elgaroyrel2}).
As pointed out in sect.~\ref{sec:tripletnumapp}, the impact on
pairing of mesonic degrees of freedom, especially through an
enhancement of the nucleon effective mass in the vicinity of
pion condensation, deserves close investigation.

In conclusion, it is important to emphasize that although the specific 
calculations performed in the present work were based on pairing matrix 
elements computed directly from bare, in-vacuum $NN$ interactions and 
on quadratic single-particle energies, many of our findings
are unaffected by these simplifications.  Indeed, the analyses that 
have been performed (leading to the universal properties discussed 
here and in ref.~\cite{univ}), as well as the calculational procedures 
that have been developed, will retain their applicability and 
essential validity when realistic in-medium inputs are employed in 
the gap equations.  

Within this in mind, further studies in the same spirit -- aimed at 
developing accurate computational procedures and uncovering universal 
properties -- are of considerable interest.  As a continuation of
the work of sect.~\ref{sec:tripletnumapp}, the separation-transformed 
BCS gap equations for pure neutron matter may be solved in the
$^3$P$_2$--$^3$F$_2$ coupled channel without resorting to the angle-average 
approximation, but still within the setting of the rapidly convergent 
small-parameter expansion, and with bare pairing matrix elements and 
unrenormalized single-particle energies.  Another project that awaits 
conclusion is a perturbative treatment of the lifting of the 
universalities and degeneracies that were established for the
uncoupled $^3$P$_2$ pairing by the analysis carried out in 
ref.~\cite{univ}.  Also, the separation analysis performed in
that paper may be extended to explore the superfluid phase
diagram of liquid $^3$He -- although this task is considerably
more demanding than in the case of $^3$P$_2$ pairing in neutron 
matter.  Finally, application of the separation approach to studies 
of $^3$S$_1$--$^3$D$_1$ pairing in symmetrical nuclear matter and 
in the expanding phase of heavy-ion collisions may produce new 
insights into these intriguing problems.

\section{Acknowledgements}
This research was supported in part by the US National Science
Foundation under Grant Nos.~PHY-9602127 and PHY-9900713 and 
by the McDonnell Center for the Space Sciences.  We thank
M.~Hjorth-Jensen, {\O}.~Elgar{\o}y, and R.~B.~Wiringa for 
stimulating discussions and helpful communications.

\section{Appendix}

In this appendix we review the application of the separation method
to $^1$S$_0$ pairing \cite{kkc} and display the first two approximants
to the gap amplitude $\Delta_F \equiv \Delta(k_F)$ within a
small-parameter expansion of the separated problem \cite{luso}.  
Having set the pattern, we then sketch the extension of the 
small-parameter expansion to $^3$P$_2$ pairing.

To provide a baseline for the separation method, it is useful to
revisit the problem of a separable pairing matrix
$V(k,k') = v_0 \phi(k) \phi(k')$
with $\phi(k_F)=1$ and $v_0\equiv V^{J=0}(k_F,k_F) \neq 0$.  The
standard gap equation in the $^1$S$_0$ channel,
\begin{equation}
\Delta(k) = - \int \frac{ V(k,k') \Delta(k') } {2\sqrt{\xi^2(k')
+ \Delta^2(k')}} {\rm d}\tau_0  \label{app1}
\end{equation}
with ${\rm d}\tau_0 = k^2 dk /2\pi^2$,
becomes
\begin{equation}
\Delta(k) = - v_0\phi(k)\int  \frac{ \phi(k')\Delta(k')  }
{2\sqrt{\xi^2(k') + \Delta^2(k')}}{\rm d}\tau_0
        \, .\label{app2}
\end{equation}
(One may note that the differential volume element ${\rm d} \tau_0$ 
entering eqs.~(\ref{app1})--(\ref{app2}) and other formulas below 
differs from the volume element ${\rm d} \tau = (2/\pi)k^2 dk$ occurring in 
integrals elsewhere in this paper.  The difference arises from
a different choice of normalization in forming the pairing matrix 
elements.  In this appendix, the S-wave pairing matrix elements are 
defined as in refs.~\cite{kkc,luso}, whereas the convention employed 
by Takatsuka and Tamagaki~\cite{ttr} for general pairing matrix elements 
is adopted in the rest of the paper.)

Since the integral on the r.h.s.\ of eq.~(\ref{app2}) is just a number,
we see that $\phi(k)$ determines the shape of the gap function, while
the integral determines its scale.  Introducing a dimensionless shape
function $\chi(k)$ such that $\Delta(k)\equiv \Delta_F \chi(k)$,
the gap equation for this problem is equivalent to
\begin{eqnarray}
\chi(k) & = & \phi(k)  \, , \\
1 & = & - v_0
	\int  \frac{\phi^2(k)}
	{2\sqrt{\xi^2(k) + \Delta_F^2 \phi^2(k)}}{\rm d}\tau_0	\, .
\end{eqnarray}
Thus, one has only to solve a nonlinear equation for the
gap amplitude rather a nonlinear singular integral equation
for the gap function.   We note that a solution of this equation
exists only when $v_F$ is negative, in which case the
monotonicity of the r.h.s.\ implies that there exists one
and only one solution.

Now we turn to the general situation  where $V(k,k')$ need not
be separable.  To gain some benefits of the separable case,
let us split the pairing interaction into a separable part and
a remainder that vanishes when either momentum variable is on
the Fermi surface:
\begin{equation}
V(k,k') = v_0\phi(k)\phi(k') + W(k,k') \, ,
\end{equation}
where
 $W(k_F,k)=W(k,k_F)\equiv 0$
and, as before, $\phi(k)=V(k,k_F)/v_F$.  Again writing
$\Delta(k) = \Delta_F \chi(k)$, the gap equation (\ref{app1})
is readily seen to be equivalent to the integral equation for
the shape function $\chi(k)$
\begin{equation}
\chi(k) + \int \frac{W(k,k')\chi(k')}
	{2\sqrt{\xi^2(k) + \Delta_F^2 \chi^2(k)}}{\rm d}\tau_0 = \phi(k)
	\, , \label{chi}
\end{equation}
together with the `algebraic' equation
\begin{equation}
1 + v_0\int \frac{\phi(k)\chi(k)}
{2\sqrt{\xi^2(k) + \Delta_F^2 \chi^2(k)}}{\rm d}\tau_0  = 0
	 \label{deltaf}
\end{equation}
for the gap amplitude $\Delta_F$ (assumed nonzero).  It is important
to note that since $W(k,k_F)$ is zero by construction, the integral
equation (\ref{chi}) has a nonsingular kernel, the log-singularity
of the original BCS formulation having been isolated in the amplitude 
equation (\ref{deltaf}).  An iterative solution of this set of equations 
converges very rapidly.  Indeed, for neutron matter at densities
corresponding to the inner-crust region of a neutron star,
a good approximation to the gap is already obtained by replacing
$\Delta_F \chi(k)$ in the denominator of (\ref{chi}) by
a small constant scale factor $\delta$ (e.g.\ $0.01 \varepsilon_F$)
solving (\ref{chi}) for $\chi(k)$ by matrix inversion, and
substituting the result into (\ref{deltaf}) to obtain the scale
factor $\Delta_F$ by (say) Newton's method.

The existence of a small parameter $d_F=\Delta_F/\varepsilon_F$
for the problem plays a key role in the success of the separation method
based on eqs.~(\ref{chi})--(\ref{deltaf}).  Consider the equation
(\ref{chi}) for the shape function $\chi(k)$. Since the Taylor expansion
of the remainder function $W(k,k')$ around $k'=k_F$ starts with
the first-order term ${\rm d} W(k,k')/{\rm d} k'|_{k'=k_F} (k' - k_F)$,
the integral over $k'$ in the symmetric vicinity
$[k_F - \delta k,k_F + \delta k]$ of this point
vanishes for sufficiently smooth functions $W(k,k')$ and $\chi(k)$,
\begin{equation}
\left.\frac{\partial^2 W(k,k')}{\partial {k'}^2}\right|_{k_F} \delta k \ll
	\left.\frac{\partial W(k,k')}{\partial k'}\right|_{k_F}
\quad \mbox{and} \quad
\frac{{\rm d} \chi(k)}{{\rm d} k} \delta k \ll 1 \, .
\end{equation}
Also, if beyond this interval we have
$\Delta_F \chi(k) \ll \xi(k)$,
or in other words if $\delta k \gg d_Fk_F$,
it is a safe approximation to replace the $\Delta_F$ term in
the denominator of (\ref{chi}) by a rather arbitrary scale factor
$\delta$. We thereby obtain a practically equivalent {\it linear}
integral equation
\begin{equation}
\chi(k) + \int \frac{W(k,k')\chi(k')}
	{2\sqrt{ \xi^2(k')+ \delta^2}}{\rm d}\tau_0 =  \phi(k) \, ,
\end{equation}
which is always soluble for well-behaved interactions.  For most
purposes, it is permissible simply to set $\delta$ equal to zero.
The nature of this ``$\delta$ approximation,'' and more generally
of expansions in the small parameter $d_F$, is developed more
fully below.

For realistic nuclear potentials, the accuracy of this approximation
in the singlet-S problem is in fact {\it much} higher than indicated by
the simple error estimate $d_F^2\ln 1/d_F$ that accompanies the above
preceding.  The reason for this heightened accuracy is that the main
contribution to the integral in (\ref{chi}) comes from a region well 
separated from the Fermi surface.  The $^1$S$_0$ pairing matrix elements 
of realistic nuclear potentials show maxima at $k\sim 3$ fm$^{-1}$.  
Hence the contribution from the Fermi surface region, already small 
because of the structure of $W(k,k')$, becomes quite negligible in 
comparison.

Based on Taylor-series expansion of the energy denominator of
eq.~(\ref{deltaf}), the gap amplitude $\Delta_F$ may be systematically
expressed as a combination of series of terms in $(d^2_F)^n\ln 1/d_F$ and
$(d^2_F)^n$ with integral $n$.  Introducing the dimensionless variable
$\epsilon (k/k_F)^2 -1$, eq.~(\ref{deltaf})
is first recast as
\begin{equation}
\frac{2}{N(0)v_F} + \int^1_{-1}	\frac{\sqrt{\epsilon+1}\phi(\epsilon)
\chi(\epsilon) {\rm d}\epsilon} {2\sqrt{\epsilon^2 + d^2_F\chi^2(\epsilon)}}
+ \int_1^{\infty}\frac{\sqrt{\epsilon+1}\phi(\epsilon)\chi(\epsilon)}
{2\sqrt{\epsilon^2 + d^2_F\chi^2(\epsilon)}} {\rm d}\epsilon
= 0 \, . \label{split}
\end{equation}
The integration over $\epsilon$ has been divided two intervals, namely
(i) the vicinity of Fermi surface $\epsilon\in[-1,1]$, where terms of
both types indicated above will appear, and (ii) the tail
$\epsilon\in[1,\infty]$, where a simple series in $d^2_F$
is generated,
\begin{equation}
\int_1^{\infty}
\frac{\sqrt{\epsilon+1}\phi(\epsilon)\chi(\epsilon)
{\rm d}\epsilon} {2\sqrt{\epsilon^2 + d^2_F\chi^2(\epsilon)}}
\simeq \int^\infty_1 \frac{\sqrt{\epsilon+1}\phi(\epsilon)\chi(\epsilon)}
{2|\epsilon|}{\rm d}\epsilon -{d^2_F\over 2}\varepsilon^2_F \int^\infty_1
\frac{\sqrt{\epsilon+1}\phi(\epsilon)\chi^3(\epsilon)}{2|\epsilon|^3}
{\rm d}\epsilon + \cdots \, .
\label{onetoinfty}
\end{equation}
Extracting the logarithmic contribution from the second term of equation
(\ref{split}), we have
\begin{eqnarray}
 \frac{2}{N(0)v_F}&+&
\ln \frac2{d_F} + \ln \frac12\left(1+\sqrt{1+d^2_F}\right) \nonumber
\\
&+&
\frac12\int^1_{-1}
	\frac{\sqrt{\epsilon+1}\phi(\epsilon)\chi(\epsilon) - 1}
	{\sqrt{\epsilon^2 + d^2_F\chi(\epsilon)^2}}
+ \frac12 \int_1^\infty {\rm d}\epsilon
	\frac{\sqrt{\epsilon+1}\phi(\epsilon)\chi(\epsilon)}
	{\sqrt{\epsilon^2 + d^2_F\chi(\epsilon)^2}}
\nonumber \\
&+& \frac12\int^1_{-1}  {\rm d}\epsilon
	\frac{\sqrt{\epsilon^2 + d^2_F} - \sqrt{\epsilon^2
        + d^2_F\chi^2(\epsilon)}}
	{\sqrt{\epsilon^2 + d^2_F}\sqrt{\epsilon^2 + d^2_F\chi^2(\epsilon)}} = 0
\, . \label{df}
\end{eqnarray}
The leading approximation to $d_F$,
\begin{equation}
d^0_F = 2 \exp{\left[\frac{2}{v_FN(0)}
+ \frac12\int^1_{-1} \frac{{\rm d}\epsilon}{|\epsilon|}
	\left(\sqrt{\epsilon+1}\phi(\epsilon)\chi(\epsilon) - 1\right)
+ \frac12\int_1^\infty \frac{ {\rm d}\epsilon}{\epsilon}
	\sqrt{\epsilon+1}\phi(\epsilon)\chi(\epsilon)
\right]} \, ,
\label{df0}
\end{equation}
is obtained by setting $d_F\equiv 0$ everywhere except within
the first logarithmic term of eq.~(\ref{df}).  The first term inside
the square brackets is the usual BCS contribution
and is clearly dominant if the interaction is weak and localized
in momentum space around the Fermi surface.  If -- as is the
case for nuclear forces -- the localization
requirement is not fulfilled, the contribution from the second term of
eq.~(\ref{df0}) must also be taken into account.   One can then
derive meaningful results even for potentials with positive $v_0$.
The approximant (\ref{df0}) is offered as a useful generalization of the
BCS weak-coupling formula to a broad class of pairing interactions.

The first correction and higher corrections to the formula (\ref{df0})
may be developed as follows.  Near the Fermi surface, the denominators
appearing in eq.~(\ref{df}) are simplified by noting that
\begin{equation}
\epsilon^2 + d_F^2\chi^2(\epsilon) = \epsilon^2 + d_F^2 +
	d_F^2 \left[\chi^2(\epsilon) - 1\right]
\end{equation}
can be expanded in powers of
$d_F^2 \left[\chi^2(\epsilon) - 1\right]/(\epsilon^2 + d_F^2) $.
This is a small quantity: for $\epsilon > d_F$ it is of order
$d^2_F/\epsilon^2$, while for $\epsilon < d_F$ we can make
the expansion
\begin{equation}
\chi(\epsilon) = 1 + \chi'_0\epsilon +
	\frac12\chi''_0\epsilon^2 + \cdots \, ,
\end{equation}
where the subscript 0 stands for $\epsilon=0$, and use
\begin{equation}
\frac{2\varepsilon d_F}{\epsilon^2 + d_F^2} \leq 1 \, .
\end{equation}
(The prime indicates the derivative with respect to $\epsilon$,
while the subscript 0 stands for $\epsilon=0$.)
Accordingly, we invoke the expansion
\begin{equation}
\frac{1}{\sqrt{\epsilon^2 + d^2_F \chi^2}} =
	\frac{1}{\sqrt{\epsilon^2 + d^2_F}}
	\left[ 1 - \frac12 d^2_F \frac{\chi^2 - 1}{\epsilon^2 + d^2_F}
        + \frac38 d^4_F \frac{{(\chi^2 - 1)}^2}{{(\epsilon^2 + d^2_F)}^2} -
         \cdots
	\right] \, . \label{chiexpansion}
\end{equation}

At this point, any smooth functions in the numerators of the integral
terms in eq.~(\ref{df}) may be expanded in Taylor series.  Since
(i) the integrations in the fourth and sixth addends on the l.h.s.\
of (\ref{df}) are performed on a symmetric interval and (ii) the
denominators contain only even powers of $\epsilon$, we need only
consider terms with even powers of $\epsilon$ in the corresponding
numerators.  For example,
\begin{equation}
\int^1_{-1}{\rm d}\epsilon
   \frac{\sqrt{\epsilon^2 + d^2_F} - \sqrt{\epsilon^2 + d^2_F\chi^2(\epsilon)}}
   {\sqrt{\epsilon^2 + d^2_F}\sqrt{\epsilon^2 + d^2_F\chi^2(\epsilon}}  =
\end{equation}
\begin{eqnarray}
&\qquad-& \frac12 d^2_F \int^1_{-1}{\rm d}\epsilon
		\frac{\chi^2 - 1}{{(\epsilon^2 + d^2_F)}^\frac32}
+ \frac38 d^4_F \int^1_{-1}{\rm d}\epsilon
	\frac{{(\chi^2 - 1)}^2}{{(\epsilon^2 + d^2_F)}^\frac52}
		- \cdots  \nonumber \\
&\qquad=&
-\frac12 d^2_F \left[(\chi'_0)^2 + \chi''_0\right]
	\left[-2 + 2 \ln\frac2{d_F}\right]
+ \frac38 d^4_F 4 (\chi'_0)^2 \left[\frac23\frac1{d^2_F} - 1\right] \cdots
\, \nonumber .
\end{eqnarray}

This step reveals a slight complication in the analysis. The integral
of the first term of the Taylor expansion of $\chi^2-1$ vanishes,
and consequently the integral of the second term of this Taylor expansion
is of the same order as the integral of the term of the
expansion (\ref{chiexpansion}) containing ${(\chi^2 - 1)}^2$
are of the same order in $d_F$.

In evaluating corrections to the result (\ref{df0}), it is convenient 
to define
\begin{equation}
I^m_n \equiv \int^1_{-1}{\rm d}\epsilon \frac{\epsilon^{2n}}
	{{(\epsilon^2 + d^2_F)}^{m+\frac12}}
\end{equation}
and
\begin{equation}
A_n \equiv (\sqrt{\epsilon+1} \phi\chi^n)''_0 \, .
\end{equation}
The first correction to $d^0_F$ is then determined by
applying recursion relations involving the $I^m_n$
(see ref.~\cite{vvkthesis}).  Eq.~(\ref{split}) is written
as
\begin{equation}
0 \simeq -\ln \frac2{d^0_F} + \frac14 d^2_F B +
	\left( 1 - \frac14 d^2_F A_3 \right) \ln \frac2{d_F} \, ,
\end{equation}
where
\begin{eqnarray}
B &=& -\int^\infty_1 \frac{\sqrt{\epsilon+1}}{\epsilon^3}
          \phi(\epsilon)\chi^3(\epsilon){\rm d}\epsilon
      -\int^1_{-1} \frac{1}{|\epsilon|^3}
         \left[ \sqrt{\epsilon+1} \phi\chi^3 -1 -
               \frac12 \epsilon^2  A_3 \right] {\rm d}\epsilon \nonumber \\
 &\quad& -\frac14 A_1 +\frac12 A_3 + \frac14 A_5 +1 \, ,
\end{eqnarray}
and the improved estimate is
\begin{equation}
d_F \simeq d^0_F \left[1 + \frac14 \left(d_F^0\right)^2
	\left(B - A_3 \ln{2\over d_F^0}\right)	\right]
\end{equation}
follows.

\begin{figure}
\begin{center}
\psfig{figure=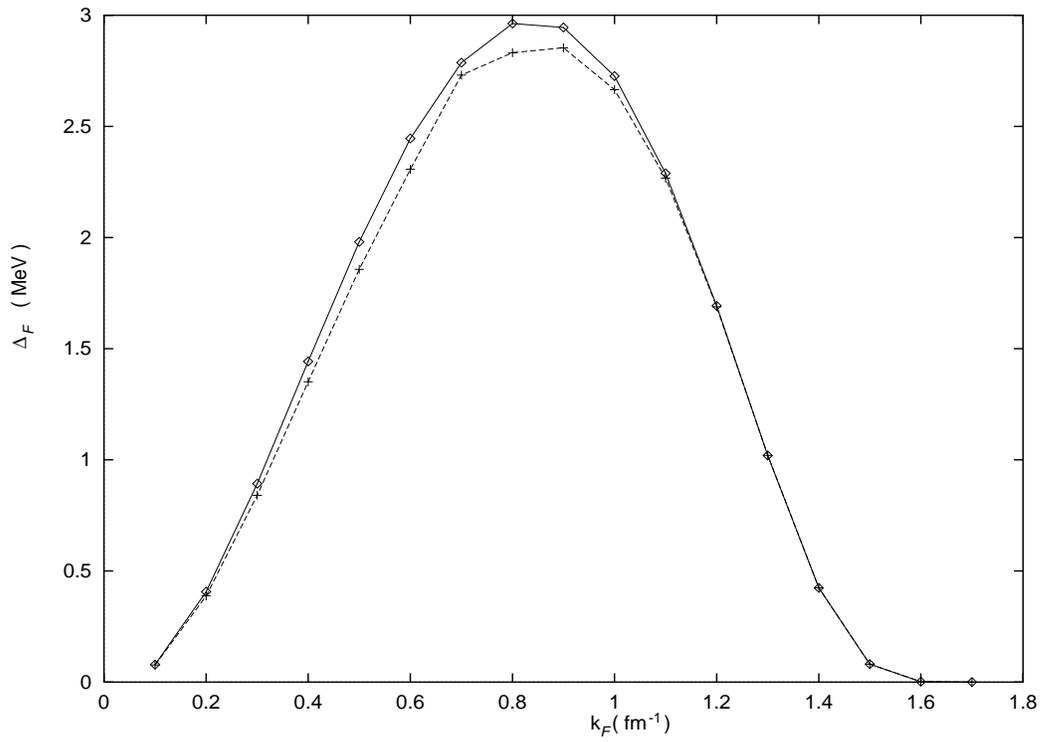,height=3.7truein,width=5.5truein,angle=-90}
\end{center}
\caption{
Superfluid energy-gap amplitude $\Delta_F$ in the $^1$S$_0$
channel, plotted as a function of Fermi momentum $k_F$ for the
Argonne $v_{18}$ potential and $m^*=1$, as calculated ``exactly''
by iteration (solid curve) and to leading approximation in
the smallness parameter $d_F$ (dashed curve).
}
\label{fig:spapprox}
\end{figure}

A calculation for the Argonne $v_{14}$ interaction illustrates the
utility and accuracy of the leading approximation (\ref{df0}) to the
energy gap $\Delta_F=\varepsilon_Fd_F$ within the small-parameter expansion.
In fig.~(\ref{fig:spapprox}), the $k_F$ dependence of $\Delta_F$ as given
by the zeroth-order approximation in $d_F$ is compared with the essentially
exact result obtained by iteration of eqs.~(\ref{chi}) and (\ref{deltaf}).
Free single-particle energies ($m^*=1$) are assumed.
The estimate (\ref{df0}) produces excellent results,
especially at densities beyond the peak of the $\Delta_F$ curve.

The same strategy of constructing approximants to the gap problem
through small-parameter expansion is applicable to pairing in higher
angular momentum states and in particular to $^3$P$_2$ pairing.
By way of illustration, consider the Type 1 ($M=\pm2$) solution of the
uncoupled-channel problem.  The transformed gap equation (\ref{mtwo}) 
associated with this case has a form similar to that for $^1$S$_0$ 
pairing, with differences arising from the anisotropy of the P-wave 
problem.  A weight factor $3(\sin^2 \theta)/2=3(1-\cos^2 \theta)/2$
appears in the integration
over angles in eq.~(\ref{mtwo}), and $3\Delta_F^2(1-\cos^2 \theta)/2$
appears in place of $\Delta_F^2$ inside the square-root energy denominator.
Doing the $k$ integral first, we can directly exploit results
and relations of sects.~\ref{sec:nocoupling} and \ref{sec:aaa} 
(specifically, the first of eqs.~(\ref{base2}) along with 
eqs.~(\ref{int10}) and (\ref{sss})).
Ignoring corrections proportional to
$d_F^2 = \Delta_F/\epsilon_F$, the only additional term that
needs to be considered, relative to the angle-average approximation,
is
\begin{equation}
\int {\rm d} {\bf n} {3\over 2}(1-\cos^2\theta)\biggl/
\ln \left[{3 \over 2}(1-\cos^2\theta) \right]\, .
\end{equation}
For the Type 2 solution (with $M=0$), the additional term has the
slightly different expression (see the last of eqs.(\ref{base2}))
\begin{equation}
\int {\rm d} {\bf n} {1 \over 2}(1+ 3\cos^2\theta)\biggl/
\ln \left[ {1\over 2}(1+ 3\cos^2\theta) \right] \,.
\end{equation}
The most salient points are that both these contributions are of
order unity relative to the dominant logarithm in the gap and both 
are negative.  We may thus conclude that the true gap will be smaller 
than that given by the angle-average approximation and that the
correction to the gap value calculated in this approximation will 
generally be small.

\newpage

\begin{table}
\caption{Energy gap $\Delta_F$ for $^3$P$_2$--$^3$F$_2$ pairing in
neutron matter at several values of the Fermi momentum, for the Argonne 
$v_{18}$ and $v_{14}$ interactions and the OPEG potential.  Free 
single-particle energies ($m^*=1$) are used, except for
the last column, where $m^*=0.78$.} 
\label{defparagcl} 
\begin{center}
\begin{tabular}{l l l l l}
\hline 
$ k_F $ &  OPEG  & $v_{14}$ & $v_{18}$ & $v_{18}(0.78)$ \\  
\hline 

1.0 &  0.0009 & 0.0020 & 0.0020 & --- \\
1.5 &  0.1369 & 0.2496 & 0.1576 & 0.0263 \\
2.0 &  0.5376 & 0.8028 & 0.4362 & 0.0947 \\
2.2 &  0.6167 & 0.8519 & 0.4235 & 0.0875 \\
2.3 &  0.6175 & 0.8182 & 0.3807 & 0.0740 \\
2.5 &  0.5419 & 0.6542 & 0.2458 & 0.0384 \\
3.0 &  0.1538 & 0.1330 & 0.0059 &  ---\\
\hline 
\end{tabular}
\end{center}
\end{table}

\newpage 

\end{document}